\pdfoutput=1 
\documentclass[lettersize,journal]{IEEEtran}
\usepackage{amsmath,amsfonts}
\usepackage{algorithmic}
\usepackage{algorithm}
\usepackage{array}
\usepackage{textcomp}
\usepackage{stfloats}
\usepackage{url}
\usepackage{verbatim}
\usepackage{graphicx}
\usepackage{cite}
\usepackage{multirow}
\usepackage{makecell}
\usepackage{pifont}
\usepackage{caption}
\usepackage{subcaption}
\usepackage[most]{tcolorbox}
\usepackage{colortbl,hhline}
\usepackage[normalem]{ulem}
\usepackage{float}
\usepackage{fontawesome}
\usepackage{subfiles}
\usepackage{booktabs}
\usepackage{bm}

\newcommand{\mxx}[1]{\textcolor{black}{#1}} 

\usepackage[colorlinks,linkcolor=blue,anchorcolor=blue,citecolor=blue]{hyperref}

\begin{document}

\title{On the Influence of Data Resampling for Deep Learning-Based Log Anomaly Detection: Insights and Recommendations}

\author{Xiaoxue Ma, Huiqi Zou, Pinjia He, Jacky Keung, Yishu Li, Xiao Yu, and Federica Sarro
\thanks{Xiaoxue Ma and Yishu Li are with the Department of Electronic Engineering and Computer Science, Hong Kong Metropolitan University, Hong Kong, China. E-mail: kxma@hkmu.edu.hk, sliy@hkmu.edu.hk.}
\thanks{Huiqi Zou is with the Department of Computer Science, Johns Hopkins University, Baltimore, United States. E-mail: hzou11@jh.edu.}
\thanks{Pinjia He is with the School of Data Science, Chinese University of Hong Kong, Shenzhen, China. E-mail: hepinjia@cuhk.edu.cn.}
\thanks{Jacky Keung is with the Department of Computer Science, City University of Hong Kong, Hong Kong, China. E-mail: jacky.keung@cityu.edu.hk.}
\thanks{Xiao Yu is with the State Key Laboratory of Blockchain and Data Security, Zhejiang University, Hangzhou, China. E-mail: xiaoyu\_cs@hotmail.com.}
\thanks{Federica Sarro is with the Department of Computer Science, University
College London, London, U.K. E-mail: f.sarro@ucl.ac.uk.}
\thanks{Xiao Yu is the corresponding author.}}


\markboth{IEEE Transactions on Software Engineering}%
{Ma \MakeLowercase{\textit{et al.}}: On the Influence of Data Resampling for Deep Learning-Based Log Anomaly Detection: Insights and Recommendations}


\maketitle

\begin{abstract}
Numerous Deep Learning (DL)-based approaches have gained attention in software Log Anomaly Detection (LAD), yet class imbalance in training data remains a challenge, with anomalies often comprising less than 1\% of datasets like Thunderbird. Existing DLLAD methods may underperform in severely imbalanced datasets. Although data resampling has proven effective in other software engineering tasks, it has not been explored in LAD.
This study aims to fill this gap by providing an in-depth analysis of the impact of diverse data resampling methods on existing DLLAD approaches from two distinct perspectives. Firstly, we assess the performance of these DLLAD approaches across four datasets with different levels of class imbalance, and we explore the impact of resampling ratios of normal to abnormal data on DLLAD approaches. Secondly, we evaluate the effectiveness of the data resampling methods when utilizing optimal resampling ratios of normal to abnormal data. Our findings indicate that oversampling methods generally outperform undersampling and hybrid sampling methods. Data resampling on raw data yields superior results compared to data resampling in the feature space. 
These improvements are attributed to the increased attention given to important tokens.
By exploring the resampling ratio of normal to abnormal data, we suggest generating more data for minority classes through oversampling while removing less data from majority classes through undersampling. 
In conclusion, our study provides valuable insights into the intricate relationship between data resampling methods and DLLAD. By addressing the challenge of class imbalance, researchers and practitioners can enhance DLLAD performance.
\end{abstract}


\section{Introduction}
\IEEEPARstart{S}{oftware}-intensive systems, which cater to a wide user base  \cite{le2022log}, are susceptible to minor issues that can lead to adverse consequences such as data corruption and performance degradation \cite{le2021log}. In this context, logs play a crucial role in system maintenance \cite{fu2013contextual,fu2014developers,jiang2017causes,zhi2019exploratory}, as they capture essential runtime information required for troubleshooting and performance monitoring \cite{he2021survey}. Consequently, there is a considerable interest in utilizing logs for anomaly detection.
Recently, many \underline{D}eep \underline{L}earning-based \underline{L}og \underline{A}nomaly \underline{D}etection (DLLAD) approaches \cite{du2017deeplog,lu2018detecting,zhang2019robust,meng2019loganomaly,le2021log,yang2021semi,ma2023semi} have been proposed to automatically identify system anomalies, showing promising results. 

In real-world scenarios in DLLAD, the proportion of normal data greatly outweighs that of abnormal data. 
For instance, consider the Thunderbird dataset in Table~\ref{table dataset}, one of the commonly used public datasets where logs are grouped into log sequences (with 20, 50, or 100 logs constituting a sequence) for data analysis. In this dataset, anomalies only account for 0.16\%--0.35\% of the total, highlighting the serious imbalance in the data distribution. 
Le et al. \cite{le2022log} have revealed that DLLAD models trained on highly imbalanced datasets exhibit low precision or recall values. Low recall leads to missed anomalies, leaving potential threats undetected, while low precision generates numerous false alarms, causing alert fatigue and resource wastage on normal logs \cite{yang2021semi,le2022log,ma2023semi}.

Despite its significance, class imbalance in DLLAD has been largely overlooked. Data resampling offers a potential solution by either generating abnormal data or removing normal data, thereby enabling the model to learn from a more balanced representation of both classes. \mxx{Previous surveys \cite{johnson2019survey, van2007experimental} indicate that no single data resampling method consistently excels across different domains. Given that log data has a unique format, often involving the grouping of log events into log sequences, which distinguishes it from other types of data, this study aims to evaluate the impact of class imbalance on DLLAD performance, identify the most effective resampling methods for DLLAD, and determine the optimal resampling ratio of normal to abnormal data.} To this end, we conduct an extensive empirical study by employing three oversampling methods (\textit{\underline{R}andom \underline{O}ver\underline{S}ampling (ROS)}, \textit{SMOTE} \cite{chawla2002smote}, and \textit{ADASYN} \cite{he2008adasyn}), three undersampling methods (\textit{\underline{R}andom \underline{U}nder\underline{S}ampling (RUS)}, \textit{NearMiss} \cite{mani2003knn}, and \textit{InstanceHardnessThreshold (IHT)} \cite{smith2014instance}), and two hybrid sampling methods (\textit{SMOTEENN} \cite{batista2004study} and \textit{SMOTETomek} \cite{batista2004study}) on DLLAD approaches (CNN \cite{lu2018detecting}, LogRobust \cite{zhang2019robust}, NeuralLog \cite{le2021log}) across four publicly available datasets using six evaluation metrics. 
The results are compared with those obtained without any resampling (\textit{NoSampling}).
Furthermore, the data resampling methods can also be categorized into resampling on raw data and resampling in the feature space.
It is important to note that many data resampling methods are designed for application only within the feature space, as they rely on distance computations. Simpler methods, like \textit{ROS} and \textit{RUS}, can be applied to both raw data (by duplicating/removing log sequences with identical texts) and feature space (by duplicating/removing sequences with the same embedding vectors).
We structure our study with the following research questions:

\textbf{RQ1: Do the existing DLLAD approaches perform well enough with varying degrees of class imbalance?}
We evaluate the performance of existing DLLAD approaches across datasets with different levels of class imbalance. \mxx{In addition, we systematically examine how class imbalance impacts these approaches while maintaining data variety.}
\textbf{Findings:} 
The performance of DLLAD approaches is quite influenced by the degree of class imbalance, with their effectiveness notably decreasing in the presence of more severe data imbalance.

\textbf{RQ2: How does the resampling ratio of normal to abnormal data affect the performance of DLLAD approaches?}
We explore how different resampling ratios of normal to abnormal data impact the DLLAD performance by using quarter-based multipliers of the original ratio of normal to abnormal log sequences.
\textbf{Findings:} The effectiveness of oversampling methods in DLLAD approaches improves when more abnormal log sequences are generated. Conversely, undersampling methods are more effective when fewer normal log sequences are removed. For hybrid sampling methods, no specific resampling ratio consistently improves DLLAD performance.

\textbf{RQ3: Does data resampling improve the effectiveness of existing DLLAD approaches?} 
We assess the effectiveness of data resampling on DLLAD approaches utilizing an optimal resampling ratio (obtained from RQ2) of normal to abnormal data.
\textbf{Findings:} 
Overall, oversampling methods demonstrate superior performance compared to undersampling and hybrid sampling methods. Remarkably, the straightforward methods applied directly to raw data outperform methods applied within the feature space.
Surprisingly, in many scenarios, certain undersampling methods (i.e., \textit{NearMiss} and \textit{IHT}), and even a hybrid sampling method \textit{SMOTEENN} aimed at mitigating data imbalance, fail to effectively enhance the performance of DLLAD approaches. 

Our study makes the following two main contributions.

1) To the best of our knowledge, we undertake the first extensive study aimed at systematically assessing the impact of data resampling methods on model performance in DLLAD. Our study encompasses a total of \mxx{5,580} experiments, wherein we employ ten data resampling methods to existing DLLAD approaches and provide a comprehensive evaluation and statistical analysis across four benchmark datasets. 
    
2) \mxx{We present findings and provide implications for researchers and practitioners in log anomaly detection. For example, we recommend utilizing \textit{ROS} on raw data for DLLAD approaches, particularly in datasets with severe class imbalance.}



\section{Background}
\subsection{Overview of DLLAD Models}

\begin{figure*}[!ht]
  \centering
  \includegraphics[width=\linewidth]{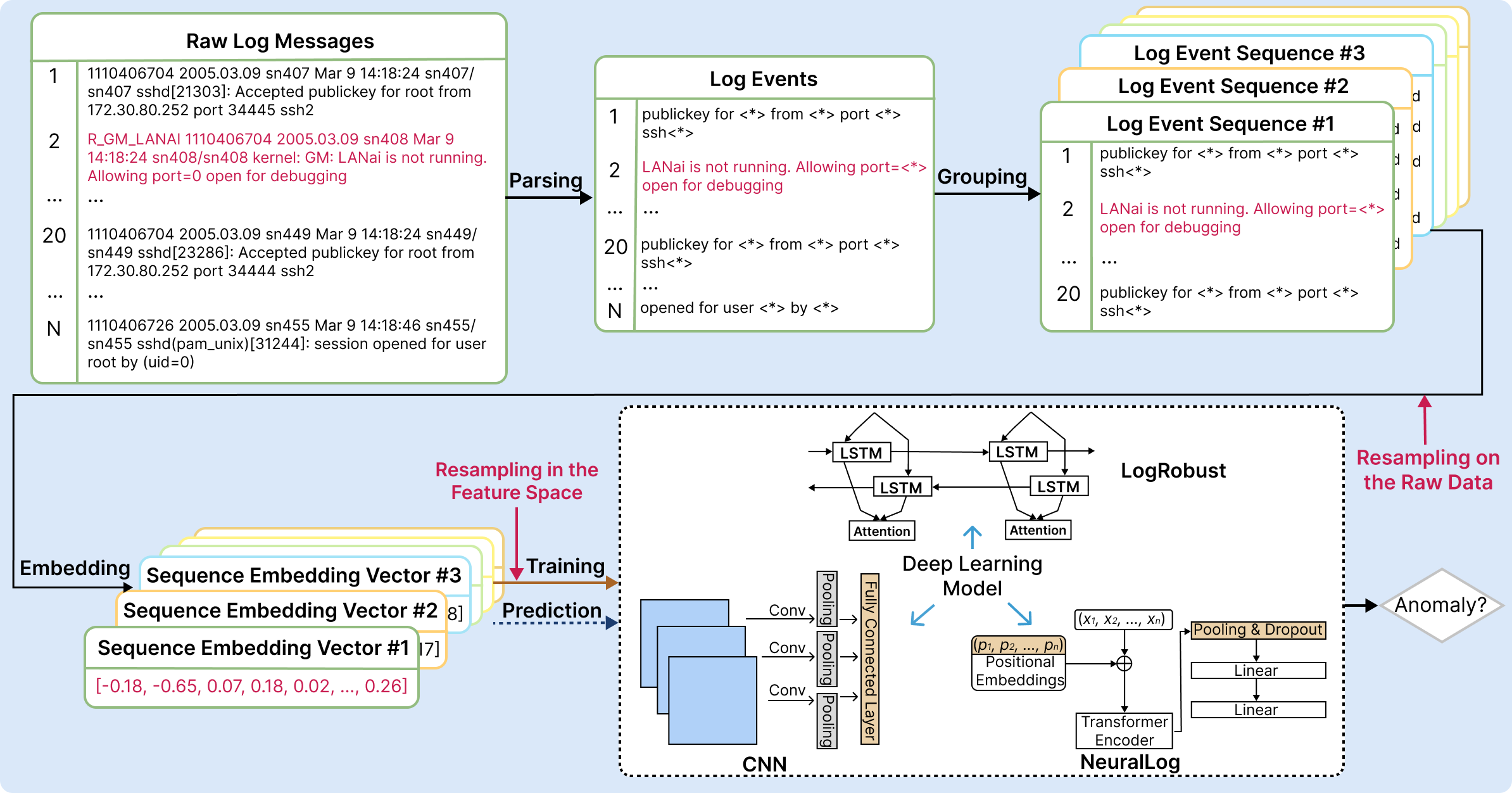}
  \caption{The common workflow of DLLAD approaches.}
   \label{fig:workflow}
\end{figure*}

The typical workflow of DLLAD approaches (shown in Figure \ref{fig:workflow}) consists of four phases: 1) log parsing, 2) log grouping, 3) log embedding, and 4) model training and prediction. To effectively extract valuable information for analysis, previous studies \cite{lin2016log, du2017deeplog, zhang2019robust, meng2019loganomaly, le2021log} convert the unstructured log messages generated during system operation into structured log events. 
Each log message comprises a header and content, where the header includes information like timestamps, typically omitted from analysis \cite{le2021log}. The log content is then segmented into constant and variable sections \cite{le2021log}. 
By replacing variable elements with a special symbol, the original log messages are converted into log events as illustrated in Figure~\ref{fig:workflow}. 
To group log events into log sequences, we adopt the fixed window strategy used in prior studies \cite{du2017deeplog, meng2019loganomaly, yang2021semi}. If an abnormal log event is part of a log sequence, the log sequence is labeled as abnormal. Conversely, if the log sequence comprises solely normal log events, it is labeled as normal. These log sequences are subsequently transformed into embedding vectors and used as input for training a classification model to predict whether a log sequence is abnormal.


\subsection{Existing DLLAD Approaches}
\label{SOTA}
Recent DL approaches for anomaly detection can be categorized into three main groups (as discussed in Section~\ref{rw}): \underline{C}onvolutional \underline{N}eural \underline{N}etwork (CNN)-based models, \underline{L}ong \underline{S}hort-\underline{T}erm \underline{M}emory-based models, and Transformer-based models. To select the most suitable DLLAD approaches, we take two factors into consideration. Firstly, we look for models that are representative of the DL models described in Section~\ref{rw}. Secondly, we aim to include models that have been recently proposed. Therefore, we choose the following models:

\textbf{CNN.} Lu et al. \cite{lu2018detecting} adopted a CNN-based model to automatically detect log anomalies.  Logs were parsed based on log keys, which were then encoded using logkey2vec. These embeddings were structured into a trainable matrix, simplifying neural network training. The model architecture comprised three convolutional layers, a dropout layer, and max-pooling layers. 

\textbf{LogRobust.} Zhang et al. \cite{zhang2019robust} employed Drain \cite{he2017drain} for log parsing and integrated the FastText \cite{joulin2016fasttext}, a pre-trained Word2vec model, with TF-IDF weights \cite{salton_term-weighting_1988} to represent log events as semantic vectors. Subsequently, these vectors are utilized as input to an attention-based \underline{Bi}-directional LSTM (Bi-LSTM) model for detecting anomalies. 

\textbf{NeuralLog.} Le et al. \cite{le2021log} preprocessed log messages without log parsing and encoded them into vector representations via a pre-trained Transformer-based language model BERT \cite{devlin2018bert}. Then, they apply a transformer encoder to classify log sequences, with the primary objective of capturing semantic information comprehensively.

\subsection{Data Resampling}
\label{resampling intro}
We make use of commonly adopted data resampling methods in the software engineering domain \cite{chakraborty2021deep, yang2023does, pelayo2012evaluating, malhotra2019empirical}. These methods are classified into three primary categories: oversampling, undersampling, and hybrid sampling.
\mxx{As depicted by the two red arrows in Figure~\ref{fig:workflow}, data resampling methods can be applied either to raw data (before embedding) or to the feature space (after embedding). Methods like random oversampling and undersampling \cite{yang2023does} are applicable to both contexts, while other methods are specific to the feature space.
Data resampling methods applied to raw data modify log sequence distribution directly in the training set to address class imbalance, which affects the number of log sequences belonging to each class and the generated embeddings. 
For example, with a set of ten log sequences (including one abnormal sequence), applying random undersampling to the raw data removes some normal sequences, leaving five normal and one abnormal sequence. Embeddings are then generated from these six resampled sequences. 
In contrast, data resampling methods applied within the feature space first generate embeddings from the original log sequences and then perform resampling based on these embeddings. For instance, if random undersampling is applied in the feature space, embeddings are created from all ten log sequences before the resampling process occurs. In the following, we provide an overview of each of the data resampling methods compared in our study.}

(1) \underline{R}andom \underline{O}ver\underline{S}ampling (ROS) randomly replicates log sequences of the minority class without generating new ones. These replicated abnormal sequences are then added to the original dataset. This method, when applied to raw data and feature space, is denoted as ROS$_R$ and ROS$_F$, respectively.

(2) \underline{R}andom \underline{U}nder\underline{S}ampling (RUS) 
randomly selects log sequences from the majority class and subsequently removes them from the original dataset. This method, when applied to raw data and feature space, is denoted as RUS$_R$ and RUS$_F$, respectively.

(3) \underline{S}ynthetic \underline{M}inority \underline{O}versampling \underline{T}echnique (SMOTE) \cite{chawla2002smote} is an oversampling method applied to the feature space. It augments the minority class by generating synthetic log sequences instead of duplications. 
This method first randomly selects log sequences from the minority class. For each selected abnormal log sequence $M_A$, one of its $k$ nearest neighbors $M_B$ are randomly chosen. The embedding vector of the synthetic log sequence $M_S$ is calculated with the formula $x_s=x_A+Random(0,1)(x_B-x_A)$, where $x_A$ and $x_B$ represent the embedding vectors of $M_A$ and $M_B$, separately. These newly generated synthetic abnormal log sequences are subsequently added to the original dataset.

(4) \underline{Ada}ptive \underline{Syn}thetic Sampling Approach \cite{he2008adasyn} (ADASYN) serves as an extension of SMOTE. Unlike SMOTE, ADASYN generates new synthetic abnormal log sequences near the class boundary instead of within the abnormal log sequences themselves. 

(5) NearMiss \cite{mani2003knn} operates as an undersampling method. It calculates the distance between two classes and randomly removes normal log sequences based on the distance. In our evaluation, we adopt NearMiss-3, which has demonstrated superior performance compared to NearMiss-1 and NearMiss-2. Specifically, NearMiss-3 selects a number of the nearest normal log sequences for each abnormal log sequence and removes them from the dataset.

(6) \underline{I}nstance \underline{H}ardness \underline{T}hreshold (IHT) \cite{smith2014instance} involves the application of a classifier to the dataset, followed by the removal of log sequences that are hard to classify.
The \underline{R}andom \underline{F}orest (RF) \cite{breiman2001random} algorithm serves as the default estimator for estimating the \underline{I}nstance \underline{H}ardness (IH) \cite{smith2014instance} of individual log sequences. 

(7) SMOTEENN \cite{batista2004study} is a hybrid sampling method that combines the oversampling method SMOTE and the undersampling method \underline{E}dited \underline{N}earest \underline{N}eighbour (ENN) \cite{wilson1972asymptotic}. 
SMOTE generates abnormal log sequences that can sometimes overlap with the majority class, making classification challenging. ENN, acting as a data cleaning method, helps 
address this issue. It removes log sequences when any or most of its closest neighbors are from a different class.

(8) SMOTETomek \cite{batista2004study} is a hybrid sampling method that shares similarities with SMOTEENN. It incorporates Tomek links \cite{tomek1976two} for data cleaning, defined by the distances between log sequences $M_i$ and $M_j$ from two classes. A pair ($M_i$, $M_j$) forms a Tomek link if no log sequence $M$ exists with $d(M_i, M) < d(M_i, M_j)$ or $d(M_j, M) < d(M_i, M_j)$. After oversampling by SMOTE, the log sequences that form Tomek links are then removed to help reduce potential noise or borderline log sequences that may affect classification performance.

\section{Study Design}
\subsection{Datasets}
\label{section dataset}
\mxx{To assess the performance of DLLAD approaches with the ten data resampling methods compared in our study, we use considered four widely used publicly available datasets (namely HDFS, BGL, Thunderbird, and Spirit) as well as a recently released integrated industrial dataset
\cite{le2021log, le2022log}. 
After careful analysis, we decided to exclude the HDFS dataset \cite{he2020loghub} from our evaluation because most existing approaches, like CNN, LogRobust, and NeuralLog, have already achieved near-optimal results on it, with F1 scores over 0.98 as reported in previous studies.
The BGL dataset \cite{oliner2007supercomputers} comprises log data from supercomputing system at Lawrence Livermore National Labs. Thunderbird and Spirit datasets \cite{oliner2007supercomputers} are acquired from two real-world supercomputers at Sandia National Labs. 
In addition, we utilize a combined annotated dataset from recent research \cite{lee2023heterogeneous}, which we refer to as Huawei. This dataset primarily consists of data from two distinct industrial cloud services within Huawei Cloud.
All these datasets consist of both normal and abnormal log messages, which have been manually identified.} 



\begin{table}[!htbp]
\caption{The statistics of the four public datasets. TB, $ws$, $Msg$, $Seq$, and $A$ are the abbreviations of the Thunderbird dataset, window sizes, Messages, Sequences, and Anomalies, respectively.}
\scriptsize
\renewcommand\arraystretch{0.8}
\label{table dataset}
\resizebox{1.0\linewidth}{!}{
\centering
\begin{tabular}{lrrrrrr}
\toprule
\multirow{2}{*}{Dataset} & \multirow{2}{*}{\# of $Msg$} & \multirow{2}{*}{$ws$} & \multicolumn{2}{c}{Training Data} & \multicolumn{2}{c}{Testing Data} \\\cmidrule(lr){4-5}\cmidrule(lr){6-7}
                            &                            &     & \# of $Seq$ & \# of $A$ & \# of $Seq$ & \# of $A$ \\\midrule
\multirow{3}{*}{BGL}        & \multirow{3}{*}{4,713,493} & 20  & 188,540         & 17,252          & 47,134          & 3,006           \\
                            &                            & 50  & 75,416          & 7,415           & 18,853          & 1,383           \\
                            &                            & 100 & 37,708          & 4,009           & 9,425           & 817             \\\midrule
\multirow{3}{*}{TB} & \multirow{3}{*}{5,000,000} & 20  & 200,000         & 328             & 50,000          & 37              \\ 
                            &                            & 50  & 79,999          & 195             & 19,999          & 29              \\
                            &                            & 100 & 39,999          & 138             & 9,999          & 23              \\\midrule
\multirow{3}{*}{Spirit}     & \multirow{3}{*}{5,000,000} & 20  & 200,000         & 8,817           & 50,000          & 290             \\
                            &                            & 50  & 79,999          & 4,275           & 19,999          & 270             \\
                            &                            & 100 & 39,999          & 2,577           & 9,999           & 250 \\\midrule  
\multirow{3}{*}{Huawei}     & \multirow{3}{*}{1,048,575} & 20  &   41,943       &    84        &    10,486      &       21       \\
                            &                            & 50  &    16,777       &      60      & 4,195          &      16        \\
                            &                            & 100 &    8,388       &      47      & 2,098           &  12\\\bottomrule  
\end{tabular}}
\end{table}

To group log events into a log sequence, a fixed window grouping strategy is commonly used  \cite{du2017deeplog, meng2019loganomaly, yang2021semi}, \mxx{where each sequence contains a fixed number $X$ of log events, with $X$ representing the window size ($ws$).}
However, choosing an appropriate $ws$ is challenging. A small $ws$ makes it difficult for log anomaly detection models to capture anomalies that span multiple log sequences \cite{le2022log}. Additionally, employing smaller $ws$ results in more log sequences containing fewer log events, ultimately leading to slower training speed. On the other hand, if $ws$ is large, log sequences may include multiple anomalies and confuse the detection scheme \cite{le2022log,liu2021lognads}. 
In the majority of prior research studies \cite{du2017deeplog, meng2019loganomaly, le2021log, yang2021semi, lee2023lanobert}, a single window size is typically employed to evaluate the proposed approaches, with $ws$=20 being the most common choice. A few studies \cite{le2022log, ma2023semi} have investigated multiple window sizes including 20, 100, and 200. In most cases, the F1 performance is found to be better at $ws$=20 and 100 compared to $ws$=200. However, there is no consistent indication of whether a window size of 20 or 100 performs better. \mxx{He et al. \cite{he2016experience} suggest that window size settings can impact the performance of supervised LAD approaches.}
Our experiment results (shown in Table~\ref{DLLAD}) also emphasize the absence of a universally optimal window size across all DLLAD datasets. For instance, LogRobust exhibits superior F1 and MCC performance on the Thunderbird dataset at $ws$=100, while achieving better performance on other datasets at $ws$=20. 
As a result, in our experiments, we consider both $ws$=20 and $ws$=100 as window sizes to account for potential variations in performance. Additionally, we introduce a $ws$ of 50 to provide a balanced perspective between the shorter and longer sequences analyzed. By including this intermediate window size, we aim to uncover a more nuanced understanding of how log sequence length impacts DLLAD performance, and whether the effects of data resampling across datasets with different window sizes are robust.

\begin{table}[!htbp]
\caption{Log variety in the datasets: E$_n$, and E$_a$ denote the number of unique normal and abnormal log events; S$_{n/a}$(20), S$_{n/a}$(50), and S$_{n/a}$(100) denote the number of unique normal/abnormal log sequences grouped by varying window sizes.}
\label{table dataset variety}
\renewcommand\arraystretch{0.8}
\resizebox{1.0\linewidth}{!}{
\centering
\begin{tabular}{l|rr|rr|rr|rr}
\toprule
Dataset &   E$_{n}$ & E$_{a}$ & S$_{n}$(20) & S$_a$(20) & S$_{n}$(50) & S$_a$(50) & S$_{n}$(100) & S$_a$(100) \\\midrule
BGL            & 1,552            & 53      & 36,262 & 3,534  & 20,074 & 2,605  & 10,889 & 1,901        \\
TB      & 3,153            & 8    & 188,722 & 241  & 92,668 & 194  & 47,937 & 155             \\
Spirit          & 95,673           & 20      & 109,817 & 8,957  & 79,103 & 4,495  & 45,384 & 2,804           \\
Huawei         & 90              & 28   & 23,635 & 105  & 20,354 & 76  & 10,424 & 59 \\\bottomrule     
\end{tabular}}
\end{table}

In Table~\ref{table dataset}, we detail the number of log sequences (\# of Sequences) for each dataset across various window sizes, as well as the count of abnormal log sequences (\# of Anomalies) within both training and test sets.
\mxx{To assess log data variety, Table~\ref{table dataset variety} presents the quantitative statistics on the variety of log events and log sequences. The BGL and Spirit datasets contain thousands of unique abnormal log sequences (i.e., log sequences that appear only once in the dataset), whereas the Thunderbird and Huawei datasets exhibit only hundreds. This analysis enables a comprehensive evaluation of DLLAD approaches across datasets with varying levels of abnormal log sequence variety.}

In Table~\ref{table sampling ratio}, we report the anomaly proportions (i.e., the proportions of abnormal sequences among all sequences) before and after employing data resampling methods. The anomaly proportions are adjusted according to the resampling ratios of normal to abnormal log sequences, which are obtained by multiplying the original ratio by quarter-based constants.
\mxx{For example, starting with an original ratio of normal to abnormal log sequences being 20:1, we apply quarter-based constants (1/4, 1/2, and 3/4) to derive new ratios of 5:1, 10:1, and 15:1, respectively. Consequently, the desired anomaly proportions are 1/6, 1/11, and 1/16, respectively.}
The datasets, as detailed in Table~\ref{table sampling ratio}, exhibit very low original anomaly proportions, ranging from 0.16\% to 10.63\%. Moreover, enlarging the window size has minimal impact on the level of class imbalance across each dataset. For example, in BGL, the anomaly proportion is 9.15\% with $ws$=20 and 10.63\% with $ws$=100.
After applying the specified resampling ratios, the anomaly proportions have shown substantial increases, such as from 9.15\% to 28.72\% (BGL dataset with $ws$=20), 9.83\% to 30.37\% (BGL dataset with $ws$=50), and 10.63\% to 33.03\% (BGL dataset with $ws$=100). 
\mxx{The hybrid sampling methods \textit{SMOTEENN} and \textit{SMOTETomek} combine oversampling and undersampling to achieve a desired anomaly proportion. \textit{SMOTE} generates synthetic samples for the minority class, while ENN or Tomek Links removes noisy or borderline samples, which may result in slight discrepancies between the final and desired anomaly proportions. The last three columns of Table~\ref{table sampling ratio} show the average anomaly proportions resulting from the application of  \textit{SMOTEENN} and \textit{SMOTETomek}.}

\begin{table*}[!htbp]
\caption{The anomaly proportions before and after over-/under-/hybrid sampling. $r$ denotes the original ratio of normal to abnormal log sequences in the training dataset, and quarter-based constants are represented as 1/4, 1/2, and 3/4. The last nine columns correspond to the anomaly proportions after data resampling.}
\label{table sampling ratio}
\renewcommand\arraystretch{0.8}
\centering
\begin{tabular}{lrr|rrrrrrrrr}
\toprule
\multirow{2}{*}{Dataset} &
  \multirow{2}{*}{$ws$} &
  \multicolumn{1}{r}{\multirow{2}{*}{\makecell[r]{Original Anomaly\\Proportion}}} &
  \multicolumn{3}{c}{\begin{tabular}[c]{@{}c@{}} Anomaly Proportion \\ After Oversampling\end{tabular}} &
  \multicolumn{3}{c}{\begin{tabular}[c]{@{}c@{}}Anomaly Proportion \\ After Undersampling\end{tabular}} &
  \multicolumn{3}{c}{\begin{tabular}[c]{@{}c@{}} Anomaly Proportion \\ After Hybrid Sampling\end{tabular}} \\ \cmidrule(lr){4-6}\cmidrule(lr){7-9}\cmidrule(lr){10-12}
 &
   &
  \multicolumn{1}{r}{} &
  \multicolumn{1}{r}{r*1/4} &
  \multicolumn{1}{r}{r*1/2} &
  \multicolumn{1}{r}{r*3/4} &
  \multicolumn{1}{r}{r*1/4} &
  \multicolumn{1}{r}{r*1/2} &
  \multicolumn{1}{r}{r*3/4} &
  r*1/4 &
  r*1/2 &
  r*3/4 \\ \midrule
\multirow{3}{*}{BGL}    & 20  & 9.15\%  & 28.72\% & 16.76\% & 11.84\% & 28.72\% & 16.77\% & 11.84\% & 27.97\% & 16.22\% & 11.42\% \\
                        & 50  & 9.83\%  & 30.37\% & 17.90\% & 12.69\% & 30.37\% & 17.90\% & 12.69\% &   30.02\%      & 17.66\%        &    12.52\%     \\
                        & 100 & 10.63\% & 32.25\% & 19.22\% & 13.69\% & 32.24\% & 19.22\% & 13.69\% & 33.03\% & 18.02\% & 11.92\% \\\midrule
\multirow{3}{*}{TB}     & 20  & 0.16\%  & 0.66\%  & 0.33\%  & 0.22\%  & 0.65\%  & 0.33\%  & 0.22\%  & 0.58\%  & 0.29\%  & 0.20\%  \\
                        & 50  & 0.24\%  & 0.98\%  & 0.49\%  & 0.33\%  & 0.97\%  & 0.49\%  & 0.32\%  & 0.85\%        &   0.43\%      &  0.32\%       \\
                        & 100 & 0.35\%  & 1.38\%  & 0.69\%  & 0.46\%  & 1.37\%  & 0.69\%  & 0.46\%  & 1.36\%  & 0.67\%  & 0.46\%  \\\midrule
\multirow{3}{*}{Spirit} & 20  & 4.41\%  & 15.58\% & 8.44\%  & 5.79\%  & 15.57\% & 8.44\%  & 5.79\%  & 15.61\% & 8.43\%  & 5.81\%  \\
                        & 50  & 5.34\%  & 18.42\% & 10.15\% & 7.00\%  & 18.42\% & 10.15\% & 7.00\%  &    17.70\%     &   9.50\%      &  6.81\%       \\
                        & 100 & 6.44\%  & 21.60\% & 12.11\% & 8.41\%  & 21.60\% & 12.11\% & 8.41\%  & 18.33\% & 10.04\% & 6.89\% \\\midrule
\multirow{3}{*}{Huawei} & 20  & 0.20\%  & 0.80\% & 0.40\%  & 0.27\%  & 0.80\% & 0.40\%  & 0.27\%  & 0.75\% & 0.40\%  & 0.28\%  \\
                        & 50  & 0.36\%  & 1.42\% & 0.71\% & 0.48\%  & 1.42\% & 0.71\% & 0.48\%  &   1.44\%     &   0.66\%      &  0.45\%       \\
                        & 100 & 0.56\%  & 2.20\% & 1.11\% & 0.75\%  & 2.20\% & 1.11\% & 0.75\%  &2.15\% & 1.10\% & 0.73\% \\\bottomrule
\end{tabular}
\end{table*}

\subsection{Evaluation}
We use four commonly used evaluation metrics Recall, Precision, Specificity, and F1-score in previous DLLAD studies \cite{yang2021semi,le2021log,le2022log,lee2023heterogeneous}. 
Given that \underline{M}atthews \underline{C}orrelation \underline{C}oefficient (MCC)  and \underline{A}rea \underline{U}nder the \underline{C}urve (AUC) are recommended for evaluating software engineering tasks with class imbalance \cite{song2018comprehensive, yao2020assessing, moussa2022use, bennin2019relative, bennin2022empirical}, we include both MCC and AUC in our evaluation to provide a comprehensive assessment of DLLAD model performance.
The commonly used four metrics originate from the confusion matrix, which describes four types of instances: TP (True Positives) represents the number of abnormal log sequences correctly predicted as anomalies, TN (True Negatives) represents the number of normal log sequences correctly predicted as normal, FP (False Positives) represents the number of normal log sequences incorrectly predicted as anomalies, and FN (False Negatives) represents the number of abnormal log sequences incorrectly predicted as normal.
The definitions of these metrics are as follows:

(1) \textit{Recall}$=\frac{TP}{TP+FN}$ represents the proportion of actual anomalies that are correctly predicted by DLLAD models out of all actual anomalies present in the testing dataset. It indicates DLLAD models' ability to capture all abnormal log sequences correctly.

(2) \textit{Precision}$=\frac{TP}{TP+FP}$ measures the proportion of predicted anomalies by DLLAD models that are actual anomalies out of all anomalies predicted by the models. It indicates the accuracy of the DLLAD models in identifying actual anomalies without falsely labeling normal log sequences as anomalies. 

(3) \textit{Specificity}$=\frac{TN}{TN+FP}$ represents the proportion of actual normal log sequences that are correctly predicted as normal by DLLAD models out of all actual normal log sequences. It indicates the ability of the DLLAD models to correctly identify normal log sequences as normal. 

(4) \textit{F1-score}$=\frac{2\times (Recall\times Precision)}{Recall+Precision}$ calculates the harmonic mean of Recall and Precision. It provides a balanced measure between Precision and Recall, giving equal weight to false positives and false negatives.

(5) \textit{MCC}$=\frac{TP\times TN-FP\times FN}{\sqrt{(TP+FP)(TP+FN)(TN+FP)(TN+FN)}}$ is a fully symmetric metric that takes into account all four values (TP, TN, FP, and FN) in the confusion matrix when calculating the correlation between ground truth and predicted values.

(6)  \textit{AUC} is a threshold-independent measure that can be calculated by assessing the area under the \underline{R}eceiver \underline{O}perating \underline{C}haracteristic (ROC) curve, which plots the true positive rate (Sensitivity) against the false positive rate (1 - Specificity) at various threshold settings. Unlike other metrics such as Precision, Recall, F1-score, and MCC, which depend on the choice of a threshold, AUC evaluates the classifier's performance across all possible threshold values, however this is not applicable in practice \cite{moussa2022use}.

To determine the statistical significance of the observed performance differences among these data resampling methods, we employ the Scott-Knott \underline{E}ffect \underline{S}ize \underline{D}ifference (ESD) test \cite{tantithamthavorn2016empirical} based on the assumptions of \underline{AN}alysis \underline{O}f  \underline{VA}riance (ANOVA). 
The Scott-Knott ESD test is a multiple comparison approach that leverages hierarchical clustering to partition these data resampling methods into distinct groups, exhibiting statistically significant differences at the predetermined significance level of 0.05 ($\alpha$=0.05).  There are no statistically significant differences between data resampling methods within the same group, but significant differences are observed between data resampling methods located in different groups.

\subsection{Research Questions}
\label{rq}
\textbf{RQ1. Do the existing DLLAD approaches perform well enough with varying degrees of class imbalance?}
\mxx{In this RQ, we aim to assess the impact of class imbalance on the effectiveness of existing DLLAD approaches.
First, we analyze the performance of DLLAD approaches across four datasets with varying window sizes, each exhibiting different levels of class imbalance.
Next, we focus on the most balanced dataset, BGL (as detailed in Table~\ref{table sampling ratio}). To systematically investigate the impact of class imbalance, 
we progressively remove duplicate abnormal log sequences until only unique abnormal sequences remain. This method preserves the variety of the log data, enabling us to assess how varying levels of class imbalance influence DLLAD performance.
}


\textbf{RQ2. How does the resampling ratio of normal to abnormal data affect the performance of DLLAD approaches?}
Due to the varying levels of class imbalance in the different datasets, it is challenging to establish a fixed resampling ratio of normal to abnormal data, such as maintaining a consistent 10:1 ratio of normal to abnormal log sequences across all datasets. Conducting an exhaustive exploration of countless potential ratios to identify an optimal resampling ratio for each dataset is not practically feasible. 
Considering the substantial data imbalance, we avoid pursuing a 1:1 ratio of normal to abnormal data during the data resampling process. This is particularly evident in the Thunderbird dataset, where the training set has a mere 0.16\% anomaly rate with a window size of 20. Applying a 1:1 ratio in such cases would result in excessive data removal during undersampling, leading to a considerable loss of information. Thus, this resampling ratio is not considered. Instead, we adopt the quarter as a foundational unit for our empirical investigations in a flexible and adaptive manner, as shown in Table~\ref{table sampling ratio}.

\textbf{RQ3. Does data resampling improve the effectiveness of existing DLLAD approaches?}
In this RQ, \mxx{we use the recommended resampling ratios of normal to abnormal data, as identified in RQ2, for the different categories of data resampling methods—oversampling, undersampling, and hybrid sampling.} Subsequently, we assess the effectiveness of these resampling methods when applied to existing DLLAD approaches.

\subsection{Implementations}
We implement the existing approaches introduced in Section~\ref{SOTA} using their respective GitHub repositories or reproduced codebases. Following previous works \cite{le2021log,le2022log}, in our dataset setup, the training set comprises the first 80\% of raw logs, while the remaining 20\% is allocated for testing. \mxx{Data resampling is applied exclusively to the training data.}
To reduce computational complexity and memory demands during data resampling operations, for NeuralLog \cite{le2021log}, we reduce the embedding dimension from 768 to 256.
For CNN \cite{lu2018detecting} and LogRobust \cite{zhang2019robust}, we utilize the implementations \cite{le2022log}, adhering to the instructions provided by the authors.
The implementation of data resampling methods is carried out using the Python toolbox \cite{lemaitre2017imbalanced} Imbalanced-learn\footnote{https://imbalanced-learn.org/stable/references/index.html}.
Our experiments encompass 36 distinct instances, resulting from the combination of 3 DLLAD approaches, 4 datasets, and 3 window sizes. For each experimental instance, we investigate 10 data resampling methods with 3 different resampling ratios of normal to abnormal data and \textit{NoSampling}. As a result, we have a total of 3 $\times$ 4 $\times$ 3 $\times$ ( 10 $\times$ 3 $+$ 1) combinations, summing up to 1,116 unique scenarios. To mitigate the variations in performance across different runs, we perform five runs for each data resampling method (including \textit{NoSampling}), culminating in a total of 5,580 experiments conducted in this study. 
These five-run results are utilized for statistical significance analysis using the Scott-Knott ESD test, which calculates the group ranking of each data resampling method across different datasets. Furthermore, the averages of the five-run results are provided in Tables~\ref{DLLAD}--\ref{neurallog} in Section \ref{rq results}.
We run our experiments on a Linux server with an Intel Xeon Silver 4210 CPU and four Nvidia GeForce RTX 3090-Ti GPUs.

\section{Results and Analysis}
\label{rq results}
\subsection{RQ1. Do the existing DLLAD approaches perform well enough with varying degrees of class imbalance?}
\label{section rq1}
\begin{table*}[!htbp]
\caption{The Recall, Precision, Specificity, F1-score, MCC, and AUC values of the three DLLAD approaches across the four datasets, each with three different window sizes (varying anomaly proportions). Bold font highlights the best performance among the three window sizes.}
\renewcommand\arraystretch{0.8}
\scriptsize
\centering
\begin{tabular}{l|l|rrr|rrr|rrr|rrr}
\toprule
                           &        & \multicolumn{3}{c|}{BGL}         & \multicolumn{3}{c|}{TB}                   & \multicolumn{3}{c}{Spirit}  & \multicolumn{3}{c}{Huawei}                     \\
Model                      & Metric & \makecell[r]{$ws$=20\\(9.15\%)}         & \makecell[r]{$ws$=50\\(9.83\%)} & \makecell[r]{$ws$=100\\(10.63\%)} & \makecell[r]{$ws$=20\\(0.16\%)}         & \makecell[r]{$ws$=50\\(0.24\%)}        & \makecell[r]{$ws$=100\\(0.35\%)} & \makecell[r]{$ws$=20\\(4.41\%)}         & \makecell[r]{$ws$=50\\(5.34\%)}         & \makecell[r]{$ws$=100\\(6.44\%)}  & \makecell[r]{$ws$=20\\(0.20\%)}         & \makecell[r]{$ws$=50\\(0.36\%)}         & \makecell[r]{$ws$=100\\(0.56\%)}     \\\midrule
\multirow{6}{*}{CNN}       & R      & \textbf{0.948}          & 0.907 & 0.916  & \textbf{0.495}          & 0.460          & 0.304  & 0.815          & 0.893          & \textbf{0.909}    &0.476  &\textbf{0.688} &0.333    \\
                           & P      & \textbf{0.985}          & 0.708 & 0.362  & \textbf{0.443}          & 0.169          & 0.213  & \textbf{0.797}          & 0.582          & 0.726     &1.000  &1.000 &0.800     \\
                           & S      & \textbf{0.999}          & 0.965 & 0.827  & \textbf{0.999}          & 0.997          & 0.997  & \textbf{0.999}          & 0.991          & 0.991       &1.000  &1.000  &1.000     \\
                           & F1     & \textbf{0.966} & 0.787 & 0.510  & \textbf{0.441} & 0.247          & 0.241  & 0.806          & 0.704          & \textbf{0.807}  &0.645  &\textbf{0.815} &0.471 \\
                           & MCC    & \textbf{0.964} & 0.779 & 0.509  & \textbf{0.454} & 0.277          & 0.247  & 0.805          & 0.716          & \textbf{0.807}  &0.690  &\textbf{0.829} &0.515 \\
                           & AUC    & \textbf{0.977}          & 0.919 & 0.903  & \textbf{0.866}          & 0.812          & 0.786  & 0.967          & \textbf{0.996}          & 0.995   &0.988  &\textbf{0.997} &0.928       \\\midrule 
\multirow{6}{*}{LogRobust} & R      & \textbf{0.949}          & 0.911 & 0.903  & \textbf{0.766}          & 0.609          & 0.449  & 0.892          & 0.910          & \textbf{0.918}      &\textbf{0.667}  &0.625 &0.250    \\
                           & P      & \textbf{0.860}          & 0.709 & 0.793  & 0.084          & \textbf{0.317}          & 0.277  & \textbf{0.973}          & 0.805          & 0.863         &1.000  &1.000 &0.750  \\
                           & S      & \textbf{0.989}          & 0.968 & 0.978  & 0.994          & \textbf{0.998}          & 0.997  & \textbf{1.000}          & 0.997          & 0.994     &1.000  &1.000 &1.000     \\
                           & F1     & \textbf{0.903} & 0.792 & 0.844  & 0.151          & \textbf{0.415} & 0.340  & \textbf{0.930} & 0.832          & 0.889     & \textbf{0.800}  & 0.769 & 0.375      \\
                           & MCC    & \textbf{0.897} & 0.783 & 0.831  & 0.252          & \textbf{0.437} & 0.350  & \textbf{0.931} & 0.854          & 0.885    & \textbf{0.816}  & 0.790 & 0.431       \\
                           & AUC    & \textbf{0.970}          & 0.960 & 0.954  & \textbf{0.915}          & 0.861          & 0.826  & 0.991          & 0.990          & \textbf{0.997}    & \textbf{0.995}  & 0.994 & 0.890       \\\midrule 
\multirow{6}{*}{NeuralLog} & R      & \textbf{0.896}          & 0.627 & 0.598  & \textbf{0.772}          & 0.730          & 0.756  & 0.899          & 0.931          & \textbf{0.938}  & \textbf{0.238}  & 0.095 & 0.150      \\
                           & P      & 0.852          & \textbf{0.872} & 0.671  & \textbf{0.758}          & 0.469          & 0.470  & \textbf{0.899}          & 0.864          & 0.800    & 1.000 & 1.000 & 1.000      \\
                           & S      & 0.989          & \textbf{0.991} & 0.878  & \textbf{1.000}         & 0.999          & 0.999  & 0.999          & 0.999          & 0.999    &1.000  &1.000 &1.000      \\
                           & F1     & \textbf{0.872} & 0.721 & 0.496  & \textbf{0.760} & 0.571          & 0.579  & 0.895          & \textbf{0.896} & 0.862    &\textbf{0.385}  &0.174 &0.261      \\
                           & MCC    & \textbf{0.864} & 0.718 & 0.511  & \textbf{0.762} & 0.585          & 0.596  & \textbf{0.897} & 0.896          & 0.865      &\textbf{0.488}  &0.308 &0.387    \\
                           & AUC    & \textbf{0.943}          & 0.809 & 0.738  & 0.856         & \textbf{0.865}        & 0.828  & 0.949          & \textbf{0.965}          & 0.964 &\textbf{0.619}  &0.548 &0.575        \\\bottomrule
\end{tabular}
\label{DLLAD}
\end{table*}

\mxx{Table \ref{DLLAD} summarizes the performance of CNN, LogRobust, and NeuralLog across the BGL, Thunderbird, Spirit, and Huawei datasets using three window sizes. When analyzing the impact of window size, we observe that a window size of 20 typically yields the best results, as highlighted in bold: 8/12 cases for recall, 10/12 for precision, 10/12 for specificity, 8/12 for F1, 9/12 for MCC, and 7/12 for AUC. Conversely, a window size of 100 achieves the best performance in only 1-3 out of 12 cases for these metrics. This suggests that smaller window sizes generally lead to better outcomes, likely because they enable DLLAD models to focus on a more concise and relevant set of log events, capturing anomaly features more effectively. In comparison, larger window sizes may introduce too many log events, making it more difficult to identify key features, which hinders the models’ ability to distinguish between abnormal and normal behaviors in log sequences.}

\begin{tcolorbox}
    [colback=yellow!10!white,colframe=red!40!black, sharp corners, leftrule={3pt}, rightrule={0pt}, toprule={0pt}, bottomrule={0pt}, left={2pt}, right={2pt}, top={3pt}, bottom={3pt}]
\textbf{Finding 1}: DLLAD approaches generally perform better when log sequences have fewer log events, i.e., smaller window sizes.
\end{tcolorbox}



\begin{figure*}[!htb]
	\centering
	\begin{subfigure}{0.24\linewidth}
		\includegraphics[width=\linewidth]{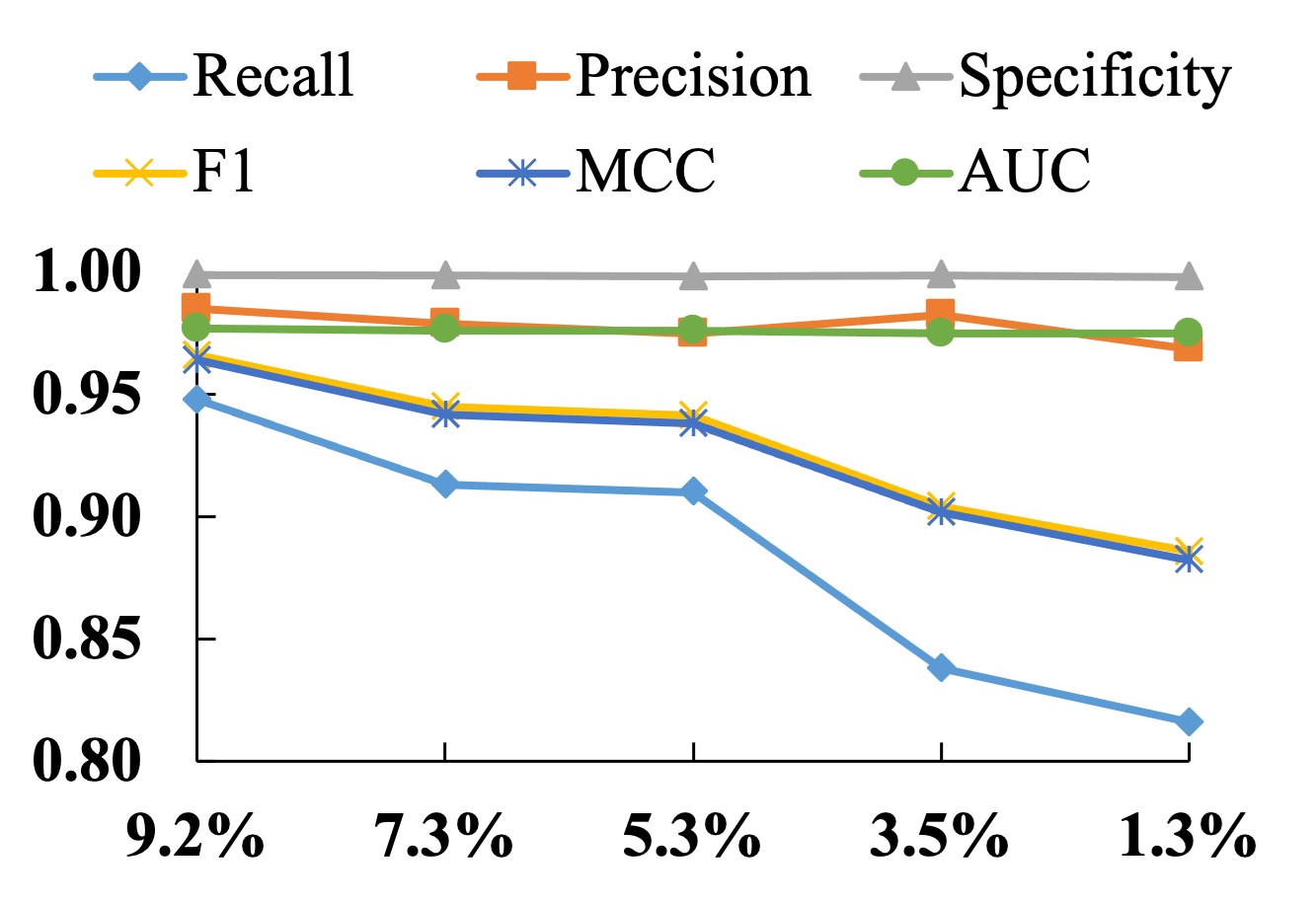}
		\caption{CNN, $ws$=20}
        \label{fig:cnn ws=20}
		
	\end{subfigure}
	\begin{subfigure}{0.24\linewidth}
		\includegraphics[width=\linewidth]{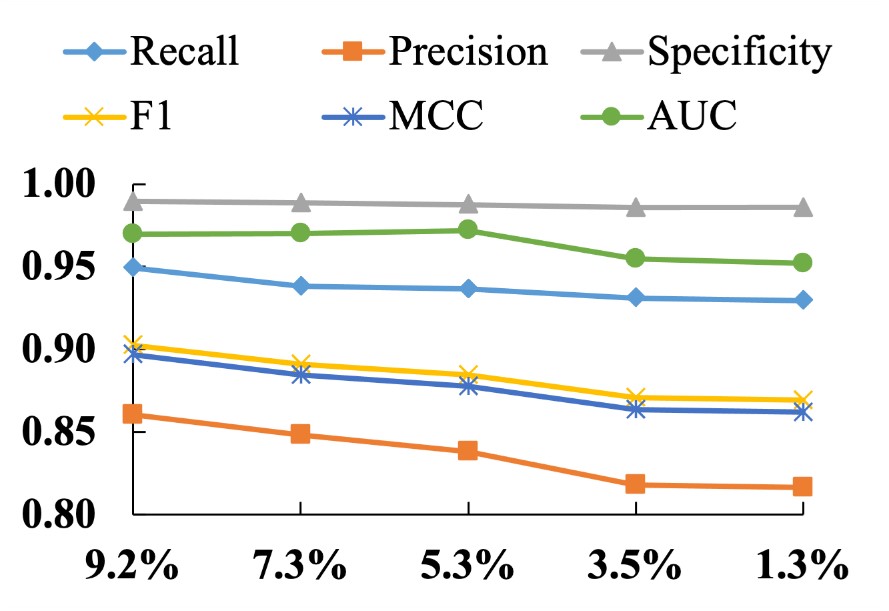}
		\caption{LogRobust, $ws$=20}
		\label{fig:lr ws=20}
	\end{subfigure}
	\begin{subfigure}{0.24\linewidth}
	        \includegraphics[width=\linewidth]{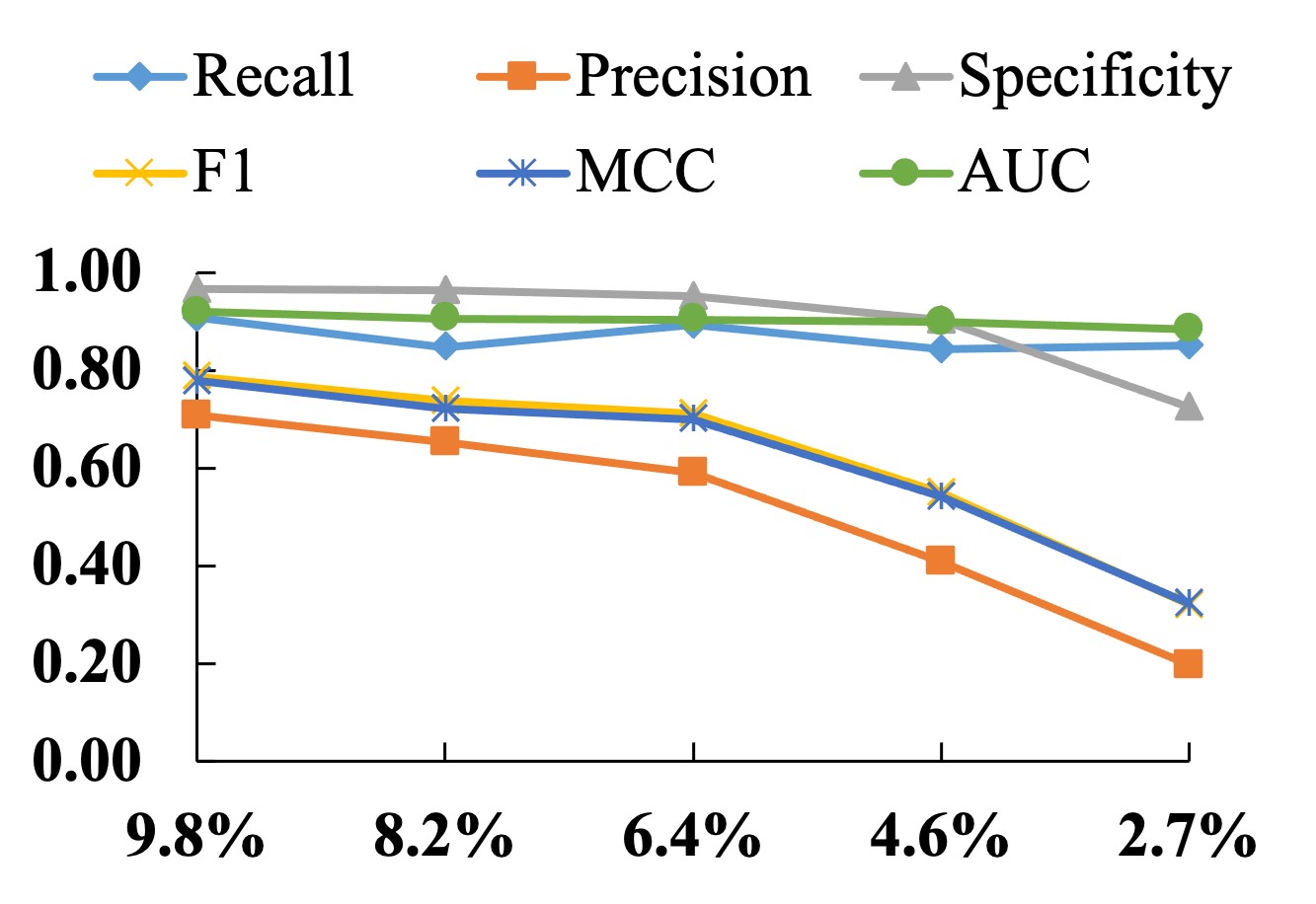}
	        \caption{CNN, $ws$=50}
	        \label{fig:cnn ws=50}
         \end{subfigure}
         	\begin{subfigure}{0.24\linewidth}
	        \includegraphics[width=\linewidth]{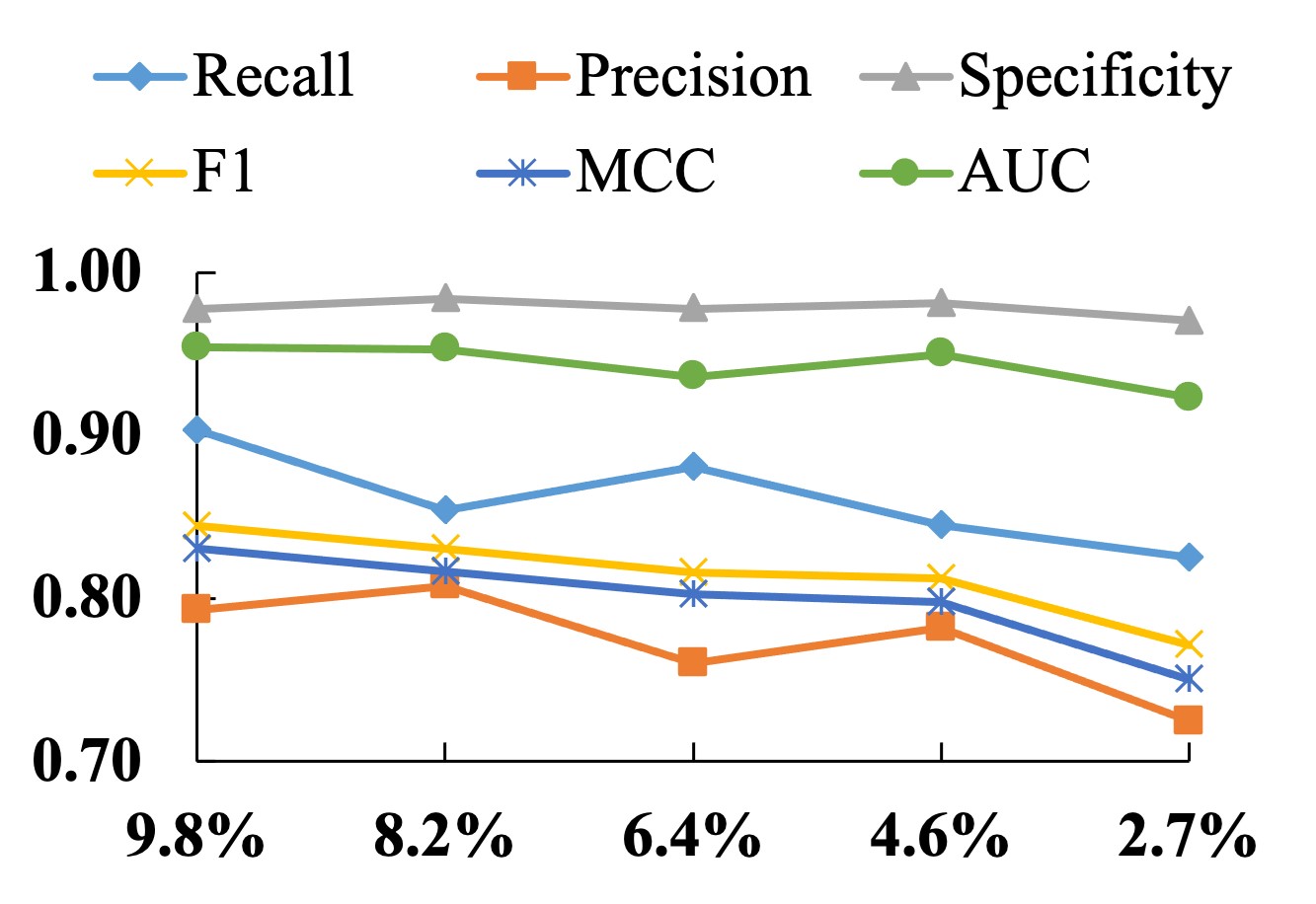}
	        \caption{LogRobust, $ws$=50}
	        \label{fig:lr ws=50}
         \end{subfigure}
	\caption{The performance variations with different anomaly proportions on the BGL dataset, while maintaining data variety.}
	\label{fig:bgl class imb}
\end{figure*}

\mxx{Regarding the severity of class imbalance (Table~\ref{table sampling ratio}), it follows the order: Thunderbird $>$ Huawei $\gg$ Spirit $>$ BGL. 
In terms of the variety of abnormal sequences (Table~\ref{table dataset variety}), the order is: Spirit $>$ BGL $\gg$ Thunderbird $>$ Huawei.
In other words, BGL and Spirit datasets are more balanced (anomaly proportion$>$4.4\%) and exhibit greater anomaly variety (thousands of unique abnormal log sequences), while Thunderbird and Huawei datasets suffer from severe class imbalance (anomaly proportion$<$0.6\%) and have less anomaly variety (up to hundreds of unique abnormal log sequences).
On the Thunderbird dataset, all three DLLAD approaches, particularly CNN and LogRobust, show poor performance, with F1 and MCC values below 0.5. Although NeuralLog achieves a better F1 score of 0.76 at $ws$=20, its overall performance remains inadequate, with F1 and MCC values below 0.6 at $ws$=50 and 100. 
A similar unsatisfactory performance of DLLAD approaches is observed in the Huawei dataset. Despite achieving nearly 100\% precision, all approaches display low recall, indicating that while most detected anomalies are true positives, many remain undetected. Notably, CNN and LogRobust outperform the more complex NeuralLog model across different window sizes. A potential reason for this is that NeuralLog's Transformer-based architecture is optimized for handling large, diverse datasets and capturing complex dependencies. However, in the Huawei dataset, which has the smallest size and least log data variety, Transformers struggle to generalize effectively, resulting in lower performance. In contrast, CNN and LogRobust are better suited for smaller datasets with less log data variety  \cite{yu2024deep}.
For the BGL and Spirit datasets, all three DLLAD approaches demonstrate strong performance, with F1, MCC, and AUC scores exceeding 0.80 at $ws$=20. This suggests that DLLAD approaches generally achieve better results when applied to more balanced datasets with greater anomaly variety.}



\mxx{To examine whether class imbalance impacts DLLAD performance while maintaining log data variety, we systematically remove varying proportions of duplicate abnormal log sequences from the most balanced BGL dataset with $ws$=50 and 100. We examine proportions of 0, 1/4, 1/2, 3/4, and all duplicates (retaining only unique sequences).
For example, if there are three distinct abnormal log sequences, each with five identical copies (totaling fifteen sequences), removing all duplicates results in three unique abnormal sequences. This method allows us to analyze how different levels of class imbalance affect DLLAD performance while maintaining log data variety.
Since NeuralLog does not perform log parsing, we focus on evaluating the other two approaches: CNN and LogRobust.
With adjustments to class imbalance through the removal of abnormal sequences, the anomaly proportions in the BGL dataset decrease from 9.2\% to 1.3\% with $ws$=50, and from 9.8\% to 2.7\% with $ws$=100. Figure~\ref{fig:bgl class imb} illustrates the performance trends for each evaluation metric corresponding to these class imbalance changes.
For CNN, as we remove increasing proportions of duplicate abnormal sequences on BGL with $ws$=20, the performance consistently declines, with decreases of 13.9\%, 1.7\%, 0.1\%, 8.3\%, 8.5\%, and 0.2\% in terms of recall, precision, specificity, F1, MCC, and AUC, respectively. Similarly, on BGL with $ws$=50, the reductions are more pronounced: 6.3\%, 72.1\%, 24.9\%, 59.3\%, 58.5\%, and 3.9\% for the same metrics.
For LogRobust, the descreases are 2.1\%, 5.1\%, 0.4\%, 3.7\%, 3.9\%, and 1.8\% for $ws$=20, and 8.7\%, 8.6\%, 0.7\%, 8.6\%, 9.7\%, and 3.2\% for $ws$=50. 
For CNN, increasing the removal of duplicate abnormal sequences from the BGL dataset results in a decline across performance metrics. With $ws$=20, recall decreases by 13.9\%, precision by 1.7\%, specificity by 0.1\%, F1 by 8.3\%, MCC by 8.5\%, and AUC by 0.2\%. When $ws$=50, these reductions become more pronounced: precision drops by 72.1\%, recall by 6.3\%, specificity by 24.9\%, F1 by 59.3\%, MCC by 58.5\%, and AUC by 3.9\%.
LogRobust shows more moderate declines. For $ws$=20, recall falls by 2.1\%, precision by 5.1\%, specificity by 0.4\%, F1 by 3.7\%, MCC by 3.9\%, and AUC by 1.8\%. For $ws$=50, the decreases are 8.7\% in recall, 8.6\% in precision, 0.7\% in specificity, 8.6\% in F1, 9.7\% in MCC, and 3.2\% in AUC.
Overall, these results highlight the negative impact of class imbalance on DLLAD performance, even when the log data variety is kept constant.}


\begin{tcolorbox}[colback=yellow!10!white,colframe=red!40!black, sharp corners, leftrule={3pt}, rightrule={0pt}, toprule={0pt}, bottomrule={0pt}, left={2pt}, right={2pt}, top={3pt}, bottom={3pt}]
  \textbf{Finding 2}: DLLAD models perform worse on datasets with more severe class imbalance. Even in datasets like BGL, which have greater log data variety, reducing the anomaly proportion (and thus increasing the severity of class imbalance) leads to a noticeable performance decline in DLLAD models.
\end{tcolorbox}


\subsection{RQ2. How does the resampling ratio of normal to abnormal data affect the performance of DLLAD approaches?}

\begin{figure}[!htb]
  \centering
	\begin{minipage}[t]{\linewidth}
		\centering
            \includegraphics[width=\linewidth]{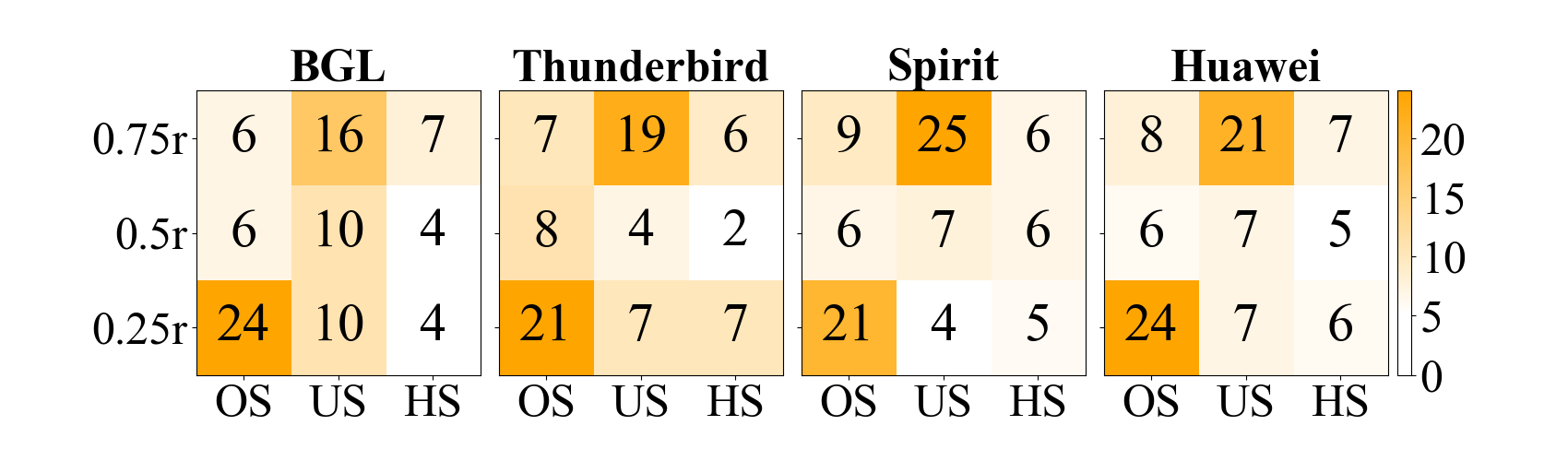}
        \subcaption{F1-score}
        \label{fig:rq2 f1}
	\end{minipage}
	\begin{minipage}[t]{\linewidth}
		\centering
		\includegraphics[width=\linewidth]{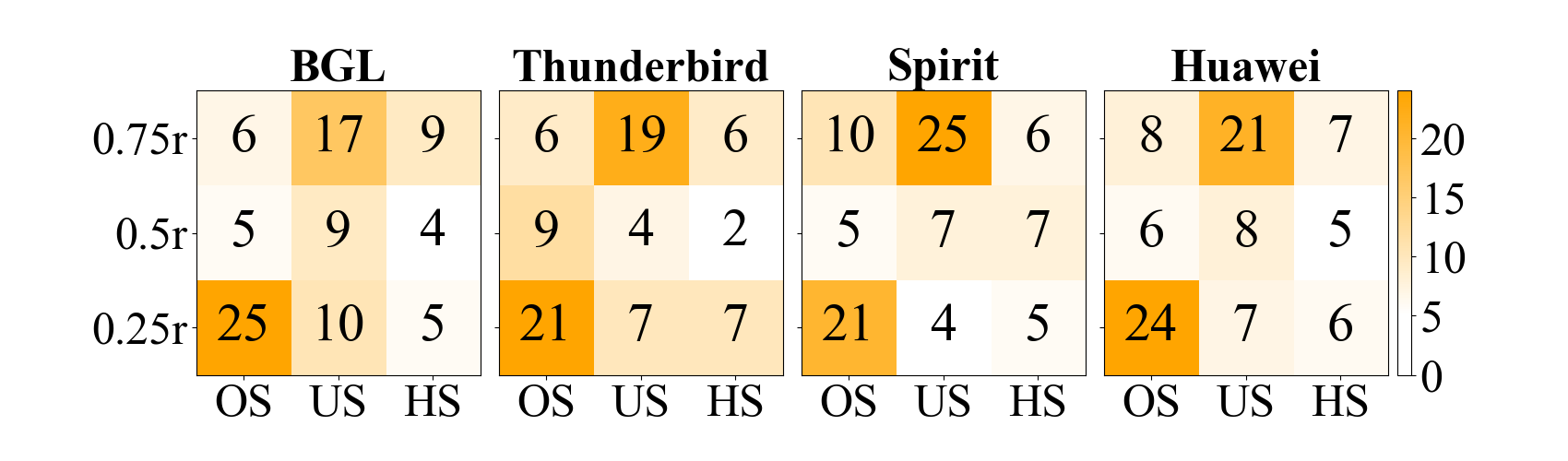}
        \subcaption{MCC}
        \label{fig:rq2 mcc}
	\end{minipage}
  \caption{The heatmaps illustrate the model performance of DLLAD approaches using different data resampling methods (OverSampling (OS), UnderSampling (US), and Hybrid Sampling (HS)) with varying resampling ratios across four datasets and three window sizes. The cells in heatmaps indicate the resampling ratio that achieves the best model performance. $r$ represents the original ratio of normal to abnormal data.}  
   \label{fig:rq2}
\end{figure}

We comprehensively evaluate the ten data resampling methods for each dataset using quarter-based resampling ratios obtained by multiplying the original ratio of normal data to abnormal data by 1/4, 1/2, and 3/4, as described in Section~\ref{section dataset}. The employed data resampling methods are categorized into three groups: \underline{O}ver\underline{S}ampling (comprising \textit{ROS$_R$}, \textit{SMOTE}, \textit{ADASYN}, and \textit{ROS$_F$}), \underline{U}nder\underline{S}ampling (encompassing \textit{NearMiss}, \textit{IHT}, \textit{RUS$_F$}, and \textit{RUS$_R$}), and \underline{H}ybrid\underline{S}ampling, represented by \textit{SMOTEENN} and \textit{SMOTETomek}. 
\mxx{Figures~\ref{fig:rq2 f1} and \ref{fig:rq2 mcc} present heatmaps that show the performance of three DLLAD approaches using these resampling methods across datasets with varying window sizes, in terms of F1-score and MCC values. We do not emphasize AUC, as the differences in AUC values across these resampling methods are minimal. Instead, we focus on two comprehensive metrics: F1-score, which balances recall and precision for detected anomalies, and MCC, a fully symmetric metric that considers all four values (TP, TN, FP, and FN) in the confusion matrix. To identify which resampling ratio optimizes performance for DLLAD approaches, we enumerate ``hits'', indicating instances where a specific resampling ratio yields the highest performance. 
For the OverSampling and UnderSampling groups, each cell in the heatmap represents 36 instances (4 oversampling/undersampling methods $\times$ 3 DLLAD approaches $\times$ 3 window sizes). For HybridSampling, each cell represents 18 instances (2 hybrid sampling methods $\times$ 3 DLLAD approaches $\times$ 3 window sizes). If the DLLAD approach achieves the same performance at two different resampling ratios, we do not count it as a ``hit''.}

\mxx{In Figures~\ref{fig:rq2 f1} and \ref{fig:rq2 mcc}, we observe that oversampling methods with a resampling ratio of one-quarter of the original normal-to-abnormal ratio consistently yield superior performance for DLLAD approaches. This means creating more abnormal sequences to achieve a target of 1/4 of the original normal-to-abnormal ratio. This trend is especially pronounced in the BGL and Huawei datasets, with 24 or 25 hits out of a total of 36 for both F1-score and MCC. For the Thunderbird and Spirit datasets, one-quarter of the original ratio results in 21 hits out of 36.
In contrast, for undersampling methods, a resampling ratio of three-quarters of the original normal log sequences achieves the best performance for DLLAD approaches. This involves reducing fewer normal sequences to reach a target of 3/4 of the original normal-to-abnormal ratio. The effect is most evident in the Spirit dataset, with 25 out of 36 hits, followed by Huawei (21 hits), Thunderbird (19 hits), and BGL (16 hits).
For hybrid sampling methods, identifying the optimal resampling ratio is challenging. In the Thunderbird, Spirit, and Huawei datasets, at least two resampling ratios show similar effectiveness in achieving the best DLLAD performance. Specifically, the most effective resampling ratios are three-quarters of the original ratio for BGL and Huawei datasets, one-quarter for Thunderbird, and one-half for Spirit. Thus, the optimal resampling ratio varies inconsistently across the four datasets.}

\begin{tcolorbox}[colback=yellow!10!white,colframe=red!40!black, sharp corners, leftrule={3pt}, rightrule={0pt}, toprule={0pt}, bottomrule={0pt}, left={2pt}, right={2pt}, top={3pt}, bottom={3pt}]
  \textbf{Finding 3}: To optimize DLLAD approaches with \textit{oversampling}, it is recommended to increase the degree of minority class amplification (i.e., generating more abnormal log sequences). Conversely, for DLLAD approaches employing \textit{undersampling}, it is advisable to reduce the degree of majority class reduction (i.e., removing fewer normal log sequences).
  
  \textbf{Finding 4}: {There is no particular preferred resampling ratio for applying \textit{hybrid sampling} methods to DLLAD approaches.}
\end{tcolorbox}


\subsection{RQ3. Does data resampling improve the effectiveness of existing DLLAD approaches?}
\label{section rq3}

Tables~\ref{cnn}--\ref{neurallog} present the results of the three DLLAD approaches with ten data resampling methods and \textit{NoSampling} on the four datasets. \mxx{In line with the RQ2 recommendation, the oversampling, undersampling, and hybrid sampling methods apply resampling ratios of one-quarter, three-quarters, and three-quarters of the original normal-to-abnormal ratio, respectively.}
We utilize three distinct colors to underscore the statistical differences among data resampling methods and \textit{NoSampling} in terms of all evaluation metrics: Recall, Precision, Specificity, F1, MCC, and AUC.
These methods are categorized into multiple groups using the Scott-Knott ESD test, often resulting in more than six groups. Darker colors indicate higher-ranked groups, with each color representing two groups (i.e., the darkest purple denotes the first and second groups, moderate purple indicates the third and fourth groups, and the lightest purple represents the fifth and sixth groups). If some data resampling methods fall outside the top six groups but significantly outperform \textit{NoSampling}, the results of these data resampling methods will be bolded.

\textbf{(1) General Insights on data resampling.} We observe that employing \textit{ROS$_R$} on all three DLLAD approaches consistently yields superior performance. Specifically, \textit{ROS$_R$} ranks within the top two groups (represented by the darkest purple color) in all \mxx{36} cases (3 DLLAD approaches $\times$ 4 datasets $\times$ 3 window sizes) in terms of at least two comprehensive metrics. \textit{SMOTE} is the second recommended data resampling method, demonstrating significantly better performance than \textit{NoSampling} in 34 out of 36 cases across all datasets in terms of at least two comprehensive metrics. 
Similarly, \textit{ADASYN} and \textit{ROS$_F$} outperforms \textit{NoSampling} in 33 out of 36 cases, followed by \textit{RUS$_R$} in 30 out of 36 cases, \textit{SMOTETomek} in 28 out of 36 cases, and \textit{RUS$_F$} in 24 out of 36 cases. 
In contrast, \textit{SMOTEENN}, \textit{NearMiss}, and \textit{IHT} exhibit poor performance, succeeding in only 14 out of 36, 11 out of 36, and 7 out of 36 cases, respectively. 
\mxx{Furthermore, we find that data resampling directly on the raw data generally statistically significantly performs better than data resampling in the feature space. For example, \textit{ROS$_R$} and \textit{RUS$_R$} perform better, equally, or worse than \textit{ROS$_F$} and \textit{RUS$_F$} in 17/18/1 and 15/16/5 out of 36 cases, respectively.
Additionally, \textit{oversampling} exhibits statistically significantly superior performance compared to \textit{undersampling} and \textit{hybrid sampling}. For instance, the best oversampling method \textit{ROS$_R$} performs better, equally, or worse than the best undersampling (\textit{RUS$_R$}) and hybrid sampling (\textit{SMOTETomek}) methods in 26/9/1 and 30/4/2 out of 36 cases, respectively.}
\begin{tcolorbox}[colback=yellow!10!white,colframe=red!40!black, sharp corners, leftrule={3pt}, rightrule={0pt}, toprule={0pt}, bottomrule={0pt}, left={2pt}, right={2pt}, top={3pt}, bottom={3pt}]
  \textbf{Finding 5}: Employing \textit{ROS$_R$}, \textit{SMOTE}, \textit{ADASYN}, and \textit{ROS$_F$} to alleviate the class imbalance can yield better results compared to \textit{NoSampling} in DLLAD.
  
  \textbf{Finding 6}: Data resampling directly on the raw data generally outperforms resampling in the feature space.
  
  \textbf{Finding 7}: \textit{Oversampling} exhibits better performance compared to \textit{undersampling} and \textit{hybrid sampling}.
\end{tcolorbox} 


\begin{table*}[!htb]
\caption{The Recall, Precision, Specificity, F1-score, MCC, and AUC values of \textbf{CNN} when employing various data resampling methods (i.e., NoSampling (NS), SMOTE (SMO), ADASYN (ADA), NearMiss (NM), InstanceHardnessThreshold (IHT), SMOTEENN (SE), SMOTETomek (ST), RandomOverSampling in the feature space (ROS$_F$), RandomUnderSampling in the feature space (RUS$_F$), RandomOverSampling applied to raw data (ROS$_R$), and RandomUnderSampling applied to raw data (RUS$_R$)) to the four datasets, each with three different window sizes. Darker cells signify superior performance, while various colors denote statistical significance among data resampling methods for each evaluation metric (determined by the Scott-Knott ESD test with a $p$-value$\textless$0.05), as observed in the subsequent tables.}
\centering
\scriptsize
\renewcommand\arraystretch{0.8}
\begin{tabular}{l|r|l|r|rr|rr|rr|rr|rr}
\toprule

Dataset &
  $ws$ &
  Metric &
  NS &
  SMO &
  ADA &
  NM &
  IHT &
  SE &
  ST &
  ROS$_F$ &
  RUS$_F$ &
  ROS$_R$ &
  RUS$_R$ \\ \midrule
 &
   &
  R &
  0.948 &
  \cellcolor{blue!22!white}0.963 &
  \cellcolor{blue!22!white}0.962 &
  \cellcolor{blue!10!white}0.956 &
  \cellcolor{blue!32!white}0.967 &
  \cellcolor{blue!22!white}0.961 &
  \cellcolor{blue!22!white}0.962 &
  \cellcolor{blue!22!white}0.962 &
  \cellcolor{blue!10!white}0.960 &
  \cellcolor{blue!22!white}0.961 &
  \cellcolor{blue!22!white}0.961 \\
 &
   &
  P &
  \cellcolor{blue!22!white}0.985 &
  \cellcolor{blue!10!white}0.978 &
  \cellcolor{blue!22!white}0.982 &
  \cellcolor{blue!32!white}0.989 &
  0.945 &
  \cellcolor{blue!32!white}0.995 &
  \cellcolor{blue!32!white}0.989 &
  \cellcolor{blue!22!white}0.984 &
  \cellcolor{blue!32!white}0.986 &
  \cellcolor{blue!22!white}0.984 &
  \cellcolor{blue!32!white}0.986 \\
 &
   &
  S &
  \cellcolor{blue!32!white}0.999 &
  \cellcolor{blue!32!white}0.999 &
  \cellcolor{blue!32!white}0.999 &
  \cellcolor{blue!32!white}0.999 &
  \cellcolor{blue!22!white}0.996 &
  \cellcolor{blue!32!white}1.000 &
  \cellcolor{blue!32!white}0.999 &
  \cellcolor{blue!32!white}0.999 &
  \cellcolor{blue!32!white}0.999 &
  \cellcolor{blue!22!white}0.992 &
  \cellcolor{blue!32!white}0.999 \\
 &
   &
  F1 &
  0.966 &
  \textbf{0.970} &
  \cellcolor{blue!10!white}0.972 &
  \cellcolor{blue!10!white}0.972 &
  0.956 &
  \cellcolor{blue!32!white}0.977 &
  \cellcolor{blue!32!white}0.975 &
  \cellcolor{blue!10!white}0.973 &
  \cellcolor{blue!22!white}0.973 &
  \cellcolor{blue!10!white}0.972 &
  \cellcolor{blue!22!white}0.974 \\
 &
   &
  MCC &
  0.964 &
  \textbf{0.968} &
  \cellcolor{blue!10!white}0.970 &
  \cellcolor{blue!10!white}0.970 &
  0.953 &
  \cellcolor{blue!32!white}0.976 &
  \cellcolor{blue!32!white}0.973 &
  \cellcolor{blue!10!white}0.971 &
  \cellcolor{blue!22!white}0.971 &
  \cellcolor{blue!22!white}0.972 &
  \cellcolor{blue!22!white}0.972 \\
\bottomrule
\end{tabular}
\label{cnn}
\end{table*}

\begin{table*}[!htb]
\caption{The Recall, Precision, Specificity, F1-score, MCC, and AUC values of \textbf{LogRobust} when employing various data resampling methods to the four datasets, each with three different window sizes.}
\centering
\scriptsize
\renewcommand\arraystretch{0.8}
\begin{tabular}{l|r|l|r|rr|rr|rr|rr|rr}
\toprule

Dataset &
  $ws$ &
  Metric &
  NS &
  SMO &
  ADA &
  NM &
  IHT &
  SE &
  ST &
  ROS$_F$ &
  RUS$_F$ &
  ROS$_R$ &
  RUS$_R$ \\ \midrule
 &
   &
  R &
  0.949 &
  \cellcolor{blue!32!white}0.958 &
  \cellcolor{blue!22!white}0.955 &
  \cellcolor{blue!10!white}0.951 &
  \cellcolor{blue!22!white}0.952 &
  \cellcolor{blue!22!white}0.952 &
  \cellcolor{blue!32!white}0.956 &
  0.938 &
  0.942 &
  \cellcolor{blue!10!white}0.951 &
  0.948 \\
 &
   &
  P &
  0.860 &
  \cellcolor{blue!10!white}0.885 &
  \cellcolor{blue!10!white}0.889 &
  \textbf{0.882} &
  0.858 &
  \cellcolor{blue!10!white}0.887 &
  \cellcolor{blue!10!white}0.889 &
  \cellcolor{blue!22!white}0.892 &
  \cellcolor{blue!22!white}0.892 &
  \cellcolor{blue!32!white}0.912 &
  \cellcolor{blue!32!white}0.904 \\
 &
   &
  S &
  0.989 &
  \cellcolor{blue!22!white}0.992 &
  \cellcolor{blue!22!white}0.992 &
  \textbf{0.991} &
  0.989 &
  \cellcolor{blue!22!white}0.992 &
  \cellcolor{blue!22!white}0.992 &
  \cellcolor{blue!22!white}0.992 &
  \cellcolor{blue!22!white}0.992 &
  \cellcolor{blue!32!white}0.994 &
  \cellcolor{blue!32!white}0.993 \\
 &
   &
  F1 &
  0.903 &
  \cellcolor{blue!22!white}0.920 &
  \cellcolor{blue!22!white}0.921 &
  \textbf{0.915} &
  0.903 &
  \cellcolor{blue!10!white}0.918 &
  \cellcolor{blue!22!white}0.921 &
  \cellcolor{blue!10!white}0.914 &
  \cellcolor{blue!10!white}0.916 &
  \cellcolor{blue!32!white}0.931 &
  \cellcolor{blue!32!white}0.925 \\
 &
   &
  MCC &
  0.897 &
  \cellcolor{blue!22!white}0.915 &
  \cellcolor{blue!22!white}0.916 &
  \textbf{0.910} &
  0.897 &
  \cellcolor{blue!10!white}0.913 &
  \cellcolor{blue!22!white}0.916 &
  \textbf{0.909} &
  \textbf{0.911} &
  \cellcolor{blue!32!white}0.926 &
  \cellcolor{blue!32!white}0.920 \\
\bottomrule
\end{tabular}
\label{logrobust}
\end{table*}

\begin{table*}[!htb]
\caption{The Recall, Precision, Specificity, F1-score, MCC, and AUC values of \textbf{NeuralLog} when employing various data resampling methods to the four datasets, each with three different window sizes.}
\centering
\scriptsize
\renewcommand\arraystretch{0.8}
\begin{tabular}{l|r|l|r|rr|rr|rr|rr|rr}
\toprule

Dataset &
  $ws$ &
  Metric &
  NS &
  SMO &
  ADA &
  NM &
  IHT &
  SE &
  ST &
  ROS$_F$ &
  RUS$_F$ &
  ROS$_R$ &
  RUS$_R$ \\ \midrule
 &
   &
R &
  0.896 &
  \cellcolor{blue!32!white}0.951 &
  \cellcolor{blue!10!white}0.929 &
  0.842 &
  0.766 &
  0.867 &
  0.875 &
  \cellcolor{blue!22!white}0.939 &
  0.897 &
  \cellcolor{blue!22!white}0.937 &
  \cellcolor{blue!10!white}0.925 \\
 &
   &
  P &
  0.852 &
  \cellcolor{blue!22!white}0.932 &
  \cellcolor{blue!10!white}0.908 &
  0.166 &
  0.818 &
  \cellcolor{blue!10!white}0.908 &
  \cellcolor{blue!22!white}0.925 &
  \textbf{0.894} &
  \cellcolor{blue!32!white}0.958 &
  \cellcolor{blue!22!white}0.923 &
  \cellcolor{blue!10!white}0.910 \\
 &
   &
  S &
  0.989 &
  \cellcolor{blue!22!white}0.995 &
  \cellcolor{blue!22!white}0.994 &
  0.691 &
  0.987 &
  \cellcolor{blue!22!white}0.994 &
  \cellcolor{blue!22!white}0.995 &
  \cellcolor{blue!10!white}0.992 &
  \cellcolor{blue!32!white}0.997 &
  \cellcolor{blue!22!white}0.995 &
  \cellcolor{blue!22!white}0.994 \\
 &
   &
  F1 &
  0.872 &
  \cellcolor{blue!32!white}0.941 &
  \cellcolor{blue!22!white}0.918 &
  0.275 &
  0.788 &
  \cellcolor{blue!10!white}0.886 &
  \cellcolor{blue!10!white}0.899 &
  \cellcolor{blue!22!white}0.915 &
  \cellcolor{blue!22!white}0.926 &
  \cellcolor{blue!32!white}0.930 &
  \cellcolor{blue!22!white}0.917 \\
 &
   &
  MCC &
  0.864 &
  \cellcolor{blue!32!white}0.937 &
  \cellcolor{blue!22!white}0.913 &
  0.282 &
  0.776 &
  \cellcolor{blue!10!white}0.880 &
  \cellcolor{blue!10!white}0.893 &
  \cellcolor{blue!22!white}0.910 &
  \cellcolor{blue!22!white}0.922 &
  \cellcolor{blue!32!white}0.925 &
  \cellcolor{blue!22!white}0.912 \\
\bottomrule
\end{tabular}
\label{neurallog}
\end{table*}

\textbf{(2) The impact of data resampling on the DLLAD approaches across datasets with varying imbalance levels.}
In datasets characterized by moderate imbalance, such as BGL and Spirit, certain data resampling methods contribute to improved performance in \mxx{all three} DLLAD approaches. 
Given the consistent results between F1 and MCC, and the minimal changes in AUC in certain cases, our analysis primarily focuses on the outcomes of MCC due to the space limitation.
\mxx{For the BGL dataset with $ws$=20, CNN (Table~\ref{cnn}), LogRobust (Table~\ref{logrobust}), and NeuralLog (Table~\ref{neurallog}) achieve high performance with \textit{NoResampling}, with MCC values of 0.964, 0.897, and 0.864, respectively. Data resampling methods, excluding \textit{IHT}, slightly improve MCC for CNN by 0.4\%--1.2\% and LogRobust by 1.3\%--3.2\%. NeuralLog shows a more significant improvement of 1.8\%--8.4\% with all resampling methods except \textit{NearMiss} and \textit{IHT}.
With $ws$=50, CNN improves by 5.5\%--14.6\% with resampling methods, except \textit{IHT}, \textit{ROS$_F$}, and \textit{RUS$_F$}. LogRobust and NeuralLog see improvements of 3.2\%--11.4\% and 3.1\%--22.4\%, respectively, with most resampling methods except \textit{NearMiss} and \textit{IHT}.
At a window size of 100, the impact of data resampling becomes more pronounced, especially in enhancing recall by reducing false positives. CNN shows a 41.8\%--78.8\% improvement with most resampling methods except \textit{IHT}. LogRobust improves by 3.2\%--11.4\% with all resampling methods, and NeuralLog by 19.0\%--69.6\%, except for \textit{NearMiss} and \textit{IHT}.
For the Spirit dataset across three window sizes, certain data resampling methods improve the performance of CNN and LogRobust, with enhancements in MCC ranging from 4.0\% to 26.0\% for CNN and 2.0\% to 8.9\% for LogRobust. These improvements occur with resampling methods, except for \textit{NearMiss}, \textit{IHT}, \textit{SMOTEENN}, \textit{SMOTETomek}, and for LogRobust, also \textit{RUS$_F$}. NeuralLog also benefits from data resampling methods, except \textit{NearMiss}, \textit{IHT}, and \textit{RUS$_F$}, with improvements of 0.9\%--11.7\%.
\textbf{Therefore, all oversampling methods (\textit{ROS$_R$}, \textit{SMOTE}, \textit{ADASYN}, and \textit{ROS$_F$}) and the undersampling method \textit{RUS$_R$} demonstrate statistically significant improvements on three DLLAD approaches in datasets with higher anomaly proportions ($\textgreater$4.4\%).}}

In the highly imbalanced Thunderbird dataset (anomaly proportion$\textless$0.4\%), \textit{ROS$_R$} and \textit{SMOTE} stand out, particularly \textit{ROS$_R$}, which shows statistically significant improvements across all cases.
For CNN (Table~\ref{cnn}), undersampling methods like \textit{RUS$_F$}, \textit{RUS$_R$}, \textit{NearMiss}, as well as the hybrid sampling method \textit{SMOTEENN}, uniformly cause the CNN to classify all log sequences as normal sequences, resulting in performance metrics (i.e., Recall, Precision, F1, and MCC) being reduced to 0. However, \textit{ROS$_R$} significantly improves \textit{NoSampling} by 48.2\%, 103.2\%, and 79.8\% for $ws$ set to 20, 50, and 100, respectively. A statistically significant difference in CNN performance is only observed with \textit{SMOTETomek} and \textit{ROS$_R$}.
LogRobust benefits from some resampling methods, except \textit{NearMiss}, \textit{IHT}, \textit{SMOTEENN}, and \textit{SMOTETomek}, with improvements ranging from 5.0\% to 121.0\% across three window sizes (shown in Table~\ref{logrobust}).
In contrast to the effects observed on CNN, certain undersampling, oversampling, and hybrid sampling methods except for \textit{ADASYN}, \textit{IHT}, and \textit{SMOTEENN}, can positively impact NeuralLog (Table~\ref{neurallog}). Particularly, data resampling applied to raw data (i.e., \textit{ROS$_R$} and \textit{RUS$_R$}) substantially enhances the MCC of \textit{NoSampling} by 6.0\% to 59.0\%.
\mxx{\textbf{Hence, data resampling methods have varied effects on three DLLAD approaches, with only \textit{ROS$_F$} being consistently beneficial on severely imbalanced datasets.}}

\mxx{In the highly imbalanced dataset Huawei (anomaly proportion$\textless$0.6\%), which has the least variety and volume compared to the public datasets (detailed in Section~\ref{section dataset}), CNN and LogRobust perform relatively better than NeuralLog (detailed in Section~\ref{section rq1}). For CNN (Table~\ref{cnn}), \textit{ADASYN}, \textit{ROS$_R$}, and \textit{RUS$_R$} improve performance by 4.5\%--25.2\% across various window sizes. For LogRobust (Table~\ref{logrobust}), \textit{ROS$_R$}, \textit{SMOTE}, \textit{ADASYN}, \textit{ROS$_F$}, and \textit{SMOTETomek} boost performance by 3.5\%--49.4\%. In contrast, NeuralLog (Table~\ref{neurallog}) shows the most significant improvements, with \textit{ROS$_R$}, \textit{SMOTE}, \textit{ADASYN}, and \textit{ROS$_F$} enhancing performance by 27.0\%--145.1\%.
The substantial performance gains observed in NeuralLog with oversampling methods suggest that generating synthetic anomalies or duplicating anomalies increases data variety and volume. As mentioned earlier, the limited size and variety of the Huawei dataset may hinder NeuralLog's complex architecture from achieving effective training. By generating anomalies, oversampling creates a richer and broader dataset, enabling the Transformer-based NeuralLog to better capture complex patterns and relationships in the log data.
In addition, we find that copying the embeddings of anomalies (\textit{ROS$_F$}) does not always enhance DLLAD model performance (e.g., CNN with $ws$=100) as effectively as copying the raw text of anomalies (\textit{ROS$_R$}). This is likely because oversampling on raw data alters the data distribution, thereby creating a wider dataset that is more effective for generating embeddings and training models.
\textbf{In short, for severely imbalanced datasets with limited data, oversampling methods, particularly those that create synthetic anomalies (e.g., \textit{ADASYN}) or duplicate raw anomalies (e.g., \textit{ROS$_R$}), can significantly enhance the performance of three DLLAD approaches.}
}

\begin{tcolorbox}
[colback=yellow!10!white,colframe=red!40!black, sharp corners, leftrule={3pt}, rightrule={0pt}, toprule={0pt}, bottomrule={0pt}, left={2pt}, right={2pt}, top={3pt}, bottom={3pt}]
   \textbf{Finding 8}: In cases of severe class imbalance (e.g., the abnormal proportion is less than 1\%), \textit{ROS$_R$} consistently enhances the effectiveness of DLLAD approaches. In cases of moderate class imbalance, \textit{ROS$_R$}, \textit{SMOTE}, \textit{ADASYN}, and \textit{ROS$_F$} exhibit effectiveness improvements across DLLAD approaches.   
\end{tcolorbox}

\mxx{Furthermore, there are other noteworthy observations from the analysis. 
Undersampling methods like \textit{NearMiss} and \textit{IHT} often perform poorly likely because reducing the majority class can lead to the loss of important information in the log data.
Similarly, the hybrid method \textit{SMOTEENN} struggles to enhance DLLAD performance. This method generates synthetic abnormal sequences and then removes log sequences considered noisy or overlapping with the majority class. However, this removal step can also discard valuable sequences that share features with both classes.
In contrast, oversampling methods, which focus on generating additional minority class sequences, effectively address class imbalance. Although these methods might introduce some noise, they generally better preserve essential information in the log data.
These observations highlight the trade-off between balancing class distributions and retaining critical data when applying data resampling methods in DLLAD.}

\begin{tcolorbox}[colback=yellow!10!white,colframe=red!40!black, sharp corners, leftrule={3pt}, rightrule={0pt}, toprule={0pt}, bottomrule={0pt}, left={2pt}, right={2pt}, top={3pt}, bottom={3pt}]
   \textbf{Finding 9}: Undersampling methods \textit{NearMiss} and \textit{IHT} are often ineffective in addressing the data imbalance problem. In addition, the hybrid method \textit{SMOTEENN} generally does not improve the performance of DLLAD approaches.
\end{tcolorbox}    

\section{Discussion}

\begin{table*}[!htb]
\caption{The frequently identified important tokens in abnormal log sequences.}
\centering
\begin{tabular}{l|l||r|r|r|r|r|r|r|r}
\hline
Model     & Resampling & network         & no & authenticated & request & error & cannot & repeated & tmreply \\

\hhline{--||--------}
CNN       & \textit{NoSampling} & .046 & .040 & .006 & .006 & .008 & .001 & .007 & .002 \\
& SMOTE & .026 & .047 & .007 & .010 & .011 & .005 & .010 & .006 \\
& ADASYN &.032 & .040 & .006 & .009 & .013 & .003 & .013 & .007 \\
& NearMiss & .009 & .040 & .008 & .007 & .009 & .005 & .011 & .006 \\
& IHT & .059 & .040 & .009 & .006 & .009 & .004 & .008 & .005 \\
& ROS$_F$ & .054 & .049 & .008 & .010 & .012 & .004 & .010 & .006 \\
& ROS$_R$ & .028 & .044 & .009 & .008 & .011 & .004 & .012 & .007 \\
          \hhline{--||--------}
LogRobust & \textit{NoSampling} & .039 & .044 & .007 & .004 & .007 & .002 & .010 & .004 \\
& SMOTE & .029 & .043 & .008 & .010 & .009 & .006 & .014 & .008 \\
& ADASYN & .034 & .047 & .009 & .009 & .010 & .013 & .013 & .009 \\
& NearMiss & .017 & .042 & .009 & .007 & .007 & .004 & .011 & .005 \\
& IHT & .045 & .051 & .010 & .007 & .006 & .004 & .012 & .007 \\
& ROS$_F$ & .040 & .050 & .007 & .009 & .008 & .015 & .011 & .010 \\
& ROS$_R$ & .031 & .054 & .008 & .009 & .009 & .006 & .014 & .009 \\
          \hhline{--||--------}
NeuralLog & \textit{NoSampling} & .021                     & .024        & .003                  & .003             & .003           & .000            & .003              & .001             \\
          & SMOTE      & .018                     & .033        & .004                  & .005             & .004           & .008            & .007              & .002             \\
          & ADASYN     & .019                     & .027        & .003                  & .004             & .006           & .003            & .004              & .002             \\
          & NearMiss   & .025                     & .029        & .005                  & .004             & .003           & .002            & .003              & .002             \\
          & IHT        & .030                     & .021        & .006                  & .003             & .003           & .003            & .003              & .002             \\
          & ROS$_F$     & .025 & .031        & .005                  & .006             & .005           & .003            & .005              & .002             \\
          & ROS$_R$     & .022 & .025        & .002                  & .004             & .004           & .003            & .004              & .003    
          \\\hline
\end{tabular}
\label{table: tokens}
\end{table*}

\subsection{Why Do(not) Data Resampling Methods Work?}
\label{why work}

In RQ3, we investigate the impact of various data resampling methods on the performance of DLLAD approaches. To enhance the interpretability of their performance, we utilize \underline{L}ocal \underline{I}nterpretable \underline{M}odel-agnostic \underline{E}xplanations (LIME) \cite{ribeiro2016should}, which is a widely-used model-agnostic explainable algorithm for explaining deep learning models in software engineering \cite{pornprasit2021pyexplainer,tantithamthavorn2021explainable,ledel2022studying,fan2020can, feichtner2020understanding,warnecke2020evaluating, lyu2021towards}. \mxx{LIME can identify the tokens associated with the highest attention weights in the neural network model. These tokens are considered highly important for the DLLAD model’s predictions.}

\mxx{Accordingly, we analyze the tokens with the highest attention weights for each abnormal log sequence in the test set and manually filter out less meaningful tokens, such as  ``from'', ``for'', ``by'', and ``via'' to focus on the highly important tokens. For a DLLAD model utilizing a data resampling method, the top ten most important tokens identified in each abnormal log sequence form the important token set \cite{steenhoek2023empirical}. 
In Table~\ref{table: tokens}, we use the Spirit dataset for demonstration, which contains the greatest variety of abnormal log sequences. We calculate the frequency of these important tokens, indicating how often each one appears in the important token set. A higher frequency implies more importance for the DLLAD model’s predictions.
The table presents the highly important tokens for three DLLAD models with the best four and worst two data resampling methods identified in RQ3.
We observe that DLLAD models frequently highlight behavior-related tokens (e.g., \texttt{repeated}, \texttt{authenticated}, and \texttt{request}) and negative tokens (e.g., \texttt{no}, \texttt{error}, and \texttt{cannot}).}

\begin{table}[!htb]
\caption{The abnormal log event within the abnormal log sequence is highlighted in bold. The top five tokens identified by LIT for LogRobust are listed below, with tokens from the abnormal log event shown in bold.} 
\centering
\footnotesize
\resizebox{0.48\textwidth}{!}{
\begin{tabular}{l}
\midrule\specialrule{0em}{0.6pt}{0.6pt}
\midrule
\textbf{Log Sequence}             \\\hline
from=\textless{}\#\textless{}*\textgreater{}\#@\#\textless{}*\textgreater{}\#\textgreater{},   size=\textless{}*\textgreater{}, nrcpt=\textless{}*\textgreater (queue active) \bm{$\times3$}                                                                                                                           \\
\begin{tabular}[c]{@{}l@{}}to=\textless{}\#\textless{}*\textgreater{}\#@\#\textless{}*\textgreater{}\#\textgreater{},   relay=none, delay=\textless{}*\textgreater{}, status=deferred (Name\\service error for name=\#\textless{}*\textgreater{}\# type=MX: Host not found, try again) \bm{$\times3$}\end{tabular} \\
mount request from \textless{}*\textgreater for   \textless{}*\textgreater \textless{}*\textgreater \bm{$\times3$}                                                                                                                                                                                                    \\
from \textless{}*\textgreater via \textless{}*\textgreater network   \textless{}*\textgreater{}.\textless{}*\textgreater{}.\textless{}*\textgreater{}/\textless{}*\textgreater{}: no free leases \bm{$\times4$}                                                                                                        \\
mount request from \textless{}*\textgreater for   \textless{}*\textgreater \textless{}*\textgreater{}                                                                                                                                                                                                      \\
from \textless{}*\textgreater{}                                                                                                                                                                                                                                                                            \\
from \textless{}*\textgreater via \textless{}*\textgreater network   \textless{}*\textgreater{}.\textless{}*\textgreater{}.\textless{}*\textgreater{}/\textless{}*\textgreater{}: no free leases                                                                                                           \\
running on \textless{}*\textgreater privileges.                                                                                                                                                                                                                                                            \\
password \textless{}*\textgreater for user \textless{}*\textgreater   (\#\textless{}*\textgreater{}\#@\#\textless{}*\textgreater{}\#).                                                                                                                                                                     \\
authentication for user \textless{}*\textgreater   accepted.                                                                                                                                                                                                                                               \\
\#\textless{}*\textgreater{}\#, coming from \textless{}*\textgreater   authenticated.                                                                                                                                                                                                                      \\
from \textless{}*\textgreater via \textless{}*\textgreater network   \textless{}*\textgreater{}.\textless{}*\textgreater{}.\textless{}*\textgreater{}/\textless{}*\textgreater{}: no free leases                                                                                                           \\
mount request from \textless{}*\textgreater for   \textless{}*\textgreater \textless{}*\textgreater{}                                                                                                                                                                                                      \\
closed for user \textless{}*\textgreater{}                                                                                                                                                                                                                                                                 \\
opened for user \textless{}*\textgreater by \textless{}*\textgreater{}                                                                                                                                                                                                                                     \\
closed for user \textless{}*\textgreater{}                                                                                                                                                                                                                                                                 \\
opened for user \textless{}*\textgreater by \textless{}*\textgreater{}                                                                                                                                                                                                                                     \\
publickey for \textless{}*\textgreater from \textless{}*\textgreater   port \textless{}*\textgreater ssh\textless{}*\textgreater \bm{$\times2$}                                                                                                                                                                        \\
repeated \textless{}*\textgreater times                                                                                                                                                                                                                                                                    \\
\textbf{cannot tm\_reply to   \textless{}*\textgreater{}.sadmin2 task \textless{}*\textgreater{}}                                                                                                                                                                                          \\
closed for user \textless{}*\textgreater{}                                                                                                                                                                                                                                                                 \\
opened for user \textless{}*\textgreater by \textless{}*\textgreater{}                                                                                                                                                                                                                                     \\
closed for user \textless{}*\textgreater{}                                                                                                                                                                                                                                                                 \\
opened for user \textless{}*\textgreater by \textless{}*\textgreater{}                                                                                                                                                                                                                                     \\
publickey for \textless{}*\textgreater from \textless{}*\textgreater   port \textless{}*\textgreater ssh\textless{}*\textgreater \bm{$\times2$}                                                                                                                                                                       \\
from \textless{}*\textgreater via \textless{}*\textgreater network   \textless{}*\textgreater{}.\textless{}*\textgreater{}.\textless{}*\textgreater{}/\textless{}*\textgreater{}: no free leases \bm{$\times8$}                                                                                                        \\
CMD (test -x /etc/pbs\_stat.py \&\&   /etc/pbs\_stat.py cron)                                           \\\midrule
\specialrule{0em}{0.6pt}{0.6pt}
\midrule

\textbf{Top Five Tokens} \\\hline
NoSampling: name, no, leases, network, closed \\
ROS$_R$: leases, no, \textbf{tm$\_$reply}, repeated, \textbf{cannot} \\
SMOTE: MX, accepted, free, repeated, \textbf{tm$\_$reply} \\
ADASYN: error, found, root, \textbf{tm$\_$reply}, \textbf{cannot} \\
ROS$_F$: service, free, network, no, \textbf{cannot} \\
NearMiss: /etc/pbs\_stat.py, root, authenticated, leases, free \\
IHT: authenticated, no, network, free, accepted\\ 
\midrule\specialrule{0em}{0.6pt}{0.6pt}
\midrule
\end{tabular}}
\label{table: token ex}
\end{table}

\mxx{In Table~\ref{table: tokens}, certain tokens, such as \texttt{cannot}, \texttt{tmreply}, and \texttt{request}, exhibit relatively low frequencies when DLLAD models are applied without data resampling. However, models utilizing the data resampling methods tend to focus more on these tokens, resulting in higher frequencies compared to those without data resampling. For instance, the token \texttt{cannot} shows a notable increase in frequency, with CNN and LogRobust models using \textit{SMOTE} having token frequencies up to 5 and 3 times higher, respectively, than their \textit{NoSampling} counterparts. Similarly, for \texttt{tmreply}, the token frequencies with CNN, LogRobust, and NeuralLog using \textit{ROS$_R$} are up to 3.5, 2.25, and 3 times higher, respectively, compared to these models without data resampling.
Conversely, tokens like \texttt{network}, which have higher frequencies in models without data resampling, show reduced frequencies when data resampling methods such as \textit{SMOTE} and \textit{ADASYN} are applied. This shift indicates that these oversampling methods enable models to prioritize more important tokens for the DLLAD model’s predictions.}

\mxx{To intuitively explain the observation, Table~\ref{table: token ex} presents an example of LIME results for an actual abnormal log sequence from the Spirit dataset, with the abnormal log event highlighted in bold. To conserve space, continuous repeated log events are denoted by $\times N$. 
The abnormal log event indicates that the system is unable to complete the expected task response. The key token \texttt{cannot} indicates an operation failure, typically signifying an abnormal event. Meanwhile, the subsequent token \texttt{tmreply} implies that the reply is not completed, suggesting a potential communication or handling issue between the system and \texttt{sadmin2}. This could lead to the task not being executed or properly concluded.
The table lists the top five tokens for LogRobust with and without data resampling. We observe that with \textit{NoSampling}, the top five tokens identified by the model do not appear in the abnormal log event, indicating the model’s failure to capture key abnormal information within the log sequence. In contrast, when using data resampling methods like \textit{ROS$_R$}, \textit{SMOTE}, \textit{ADASYN}, and \textit{ROS$_F$}, the model effectively identifies key tokens within the abnormal log event. Notably, \texttt{cannot} and \texttt{tmreply} rank among the top five tokens when employing the \textit{ROS$_R$} and \textit{ADASYN}.
These data resampling methods enable DLLAD models to focus more on the important features of abnormal log events, assigning greater weights to these important tokens and allowing the models to more accurately detect abnormal log sequences.}

\begin{table}[!htb]
\caption{The average data resampling time ($T_{resample}$) and model training time ($T_{train}$) for NeuralLog on four datasets with $ws$=20.}
\label{table effi}
\centering
\renewcommand\arraystretch{0.8}
\begin{tabular}{l|r|r|r}
\toprule
Resampling & $T_{resample}$ & $T_{train}$           & $T_{resample+train}$  \\\midrule
\textit{NoSampling} & -          & 18m21s          & 18m21s          \\ 
ROS$_R$      & \textless1ms       & 19m12s          & 19m12s          \\
RUS$_R$      & \textless1ms       & 14m10s & 14m10s \\
ROS$_F$      & 4m12s    & 23m48s          & 28m             \\
RUS$_F$      & 2m31s    & 14m16s & 16m47s \\
SMOTE      & 4m24s    & 23m33s          & 27m57s          \\
ADASYN        & 6m32s    & 23m38s          & 30m10s          \\
NearMiss         & 1m59s    & 16m4s  & 18m3s  \\
IHT        & 89m16s   & 19m11s          & 108m27s         \\
SMOTEENN         & 30m46s   & 18m55s          & 49m41s          \\
SMOTETomek        & 27m8s    & 21m10s          & 48m18s   
\\\bottomrule
\end{tabular}
\end{table}

\subsection{Efficiency of Data Resampling and Model Training}
\label{sec: effi}
\mxx{In Section~\ref{rq results}, we examine the effectiveness of data resampling methods applied to DLLAD approaches. Additionally, this section evaluates the efficiency of these approaches and their impact on model training time. For this analysis, we use NeuralLog, a model that typically requires more training time, applied across the four datasets with a window size of 20. The average data resampling time ($T_{resample}$) and model training time ($T_{train}$) are presented in Table~\ref{table effi}.}

\mxx{Data resampling on raw data methods, such as \textit{ROS$_R$} and \textit{RUS$_R$}, is extremely efficient, taking less than 1 millisecond, making it negligible. In contrast, data resampling methods applied in the feature space, including \textit{ROS$_F$}, \textit{RUS$_F$}, \textit{SMOTE}, \textit{ADASYN}, and \textit{NearMiss}, take up to 7 minutes. More complex methods like \textit{IHT}, \textit{SMOTEENN}, and \textit{SMOTETomek} exhibit much longer processing times, ranging from 27 to 89 minutes. Regarding model training time, undersampling methods generally lead to faster training for NeuralLog. For instance, \textit{RUS$_R$} and \textit{RUS$_F$} reduce training time by approximately 4 minutes compared to \textit{NoSampling}. In contrast, oversampling and hybrid sampling methods  increase training time, but the increase remains under 6 minutes compared to \textit{NoSampling}.}

\subsection{Implications of Our Findings}
\label{sec: impli}
\mxx{Sections~\ref{rq results} and \ref{sec: effi} offer a comprehensive analysis of the impact of various data resampling methods on the effectiveness and efficiency of DLLAD approaches. Based on the insights, we outline several practical implications for practitioners.}

\textbf{(1) Effectiveness prioritization:}
\mxx{For non-urgently training DLLAD models to detect log anomalies, practitioners prioritize high accuracy in predicted results. 
Utilizing \textit{ROS$_R$} generally enhances DLLAD model performance. Notably, in highly imbalanced datasets like Thunderbird, the performance of three DLLAD models improves consistently when employing \textit{ROS$_R$}.
In the Huawei dataset—also highly imbalanced but characterized by lower data volume and less variety—certain oversampling methods (e.g., \textit{ROS$_R$} and \textit{ADASYN}) that duplicate or create abnormal log sequences are beneficial for improving DLLAD performance. Therefore, we recommend using \textit{ROS$_R$} with the DLLAD model for effective predictions.
}

\textbf{(2) Efficiency prioritization:} 
\mxx{For scenarios requiring efficient training of DLLAD models to quickly identify log anomalies, CNN and LogRobust are preferable due to their shorter training times. Furthermore, using \textit{ROS} and \textit{RUS} methods on raw data for duplication and removal before embedding takes considerably less time than data resampling methods in feature spaces, as indicated in Table~\ref{table effi}. These methods also perform well, achieving top-1 and top-5 ranks among the ten data resampling methods evaluated (detailed in Section~\ref{section rq3}).
Based on our experimental findings, LogRobust outperforms CNN overall. Therefore, employing \textit{ROS$_R$} and \textit{RUS$_R$} with LogRobust is recommended for efficient prediction.}

\mxx{\textbf{(3) Further enhancement:} 
As discussed in Section~\ref{why work}, our analysis identifies certain important tokens that contribute to DLLAD, such as \texttt{cannot} and \texttt{tmreply} in the Spirit dataset. When employing data resampling methods like \textit{SMOTE}, \textit{ADASYN}, and \textit{ROS$_R$}, these tokens receive increased attention weights, thereby aiding DLLAD models in better-predicting log anomalies.
Building on these insights, future research may be able to develop enhanced data oversampling methods that focus on generating or duplicating abnormal log sequences containing these important tokens. By prioritizing log sequences with highly important tokens, these methods aim to improve the training process, enabling the model to learn more effectively from relevant abnormal features. This targeted data resampling strategy balances the dataset, thereby amplifying the presence of key abnormal log features and ultimately leading to improved anomaly detection performance.}


\subsection{Threats to Validity}
\textbf{(1) Limited models and datasets}. 
One potential concern pertains to our selection of DLLAD approaches, we adapted existing supervised approaches for our empirical investigation. This choice is motivated by several factors:
Semi-supervised approaches leverage only a fraction of normal logs, whereas unsupervised methods assume datasets lack labels, diverging from our fully supervised data scenario. Furthermore, some unsupervised and semi-supervised approaches share a model structure similar to NeuralLog, as observed in LAnoBERT \cite{lee2023lanobert} and Hades \cite{lee2023heterogeneous}. Previous empirical studies \cite{le2021log,le2022log} have consistently shown inferior performance of unsupervised and semi-supervised approaches compared to supervised ones. Hence, we deliberately include the latest supervised approaches as our evaluated DLLAD approaches.
Another concern arises from the limited availability of datasets. \mxx{Currently, only a few public datasets are available (e.g., HDFS, BGL, Thunderbird, and Spirit). To enhance our study, we include the recently released Huawei dataset. These datasets exhibit different levels of class imbalance and data variety, allowing our empirical study to generalize its findings more effectively.}


\textbf{(2) Hyperparameter settings.} 
The selection of an appropriate resampling ratio of normal to abnormal data constitutes a threat when assessing the effectiveness of data resampling methods. 
Considering numerous data resampling methods and the impracticality of evaluating all of them across diverse datasets with exhaustive resampling ratio variations, it is crucial to systematically choose the resampling ratio. To address this challenge, we introduce a standardized quarter-based unit for a consistent benchmark across various data resampling methods. On this basis, the general conclusions obtained aid researchers in narrowing their focus for subsequent study phases.

\textbf{(3) Generalizability.}
\mxx{We intentionally select ten data resampling methods from three distinct categories and systematically apply them to three representative DLLAD approaches across four publicly available datasets. 
Additionally, Lyu et al. \cite{lyu2021towards} highlight that the randomness introduced by data resampling (i.e., undersampling the majority class) can result in internal inconsistencies in model interpretations.
To alleviate this issue, we average the results over five runs to reduce the impact of randomness and internal inconsistencies.
Our objective does not compare the effectiveness of those DLLAD approaches but rather focuses on evaluating the capabilities of different data resampling methods applied to those approaches. As such, our findings aim to highlight general trends in how data resampling affects different DLLAD approaches, with the ultimate goal of offering valuable insights to inform future research endeavors.}

\section{Related Work}
\label{rw}
\subsection{Deep Learning-Based Log Anomaly Detection}
\mxx{Since the work by He et al. \cite{he2016experience} (which evaluated six machine learning-based anomaly detection approaches and compared their accuracy and efficiency on two representative production log datasets), numerous DLLAD models have emerged.} These DLLAD models typically fall into several categories, including CNN-based, LSTM-based, and Transformer-based models.
Lu et al. \cite{lu2018detecting} employed CNN with three filters to extract local semantic information from log data. 
Zhang et al. \cite{zhang2016automated}, Du et al. \cite{du2017deeplog}, and Meng et al. \cite{meng2019loganomaly} adopted LSTM \cite{hochreiter1997long} to capture long-term dependencies in log sequences and learn log patterns for predicting the next log. Vinayakumar et al. \cite{vinayakumar2017long} employed a stacked-LSTM model to learn temporal patterns using sparse representations. Zhang et al. \cite{zhang2019robust} proposed the LogRobust method that integrated the attention mechanism with a Bi-LSTM model, enabling comprehensive sequence information capture in both directions. Li et al. \cite{li_swisslog_2020} employed a unified attention-based Bi-LSTM model to learn the patterns for sequential anomaly detection.
Le et al. \cite{le2021log} introduced NeuralLog, utilizing BERT for embedding representation and a Transformer encoder for log anomaly detection classification.  
To the best of our knowledge, only the work by Le et al. \cite{le2022log} has highlighted that DLLAD models trained on highly imbalanced datasets exhibit low precision or recall values. Yet, there is currently no research exploring whether data resampling methods can mitigate class imbalance issues and improve DLLAD model performance. 
Subsequent work, such as unsupervised approach LAnoBERT \cite{lee2023lanobert}, and semi-supervised approaches like PLELog \cite{yang2021semi}, AdaLog \cite{ma2023semi}, and Hades \cite{lee2023heterogeneous}, have emerged to alleviate the potential difficulty of acquiring large labeled log datasets for training supervised learning models in log anomaly detection.
However, we chose to focus on CNN \cite{lu2018detecting}, LogRobust \cite{zhang2019robust}, and NeuralLog \cite{le2021log} as the DLLAD models in our study, instead of these recently proposed semi-supervised or unsupervised methods. Our rationale behind this decision is based on the following reasons. 
(1) Different data scenarios: semi-supervised approaches use only a portion of the normal logs, and unsupervised approaches assume datasets have no labels, which differs from our fully supervised data hypothesis scenario. 
(2) Similar model structure: some unsupervised and semi-supervised approaches share a model structure similar to NeuralLog \cite{le2021log}.  For example, LAnoBERT \cite{lee2023lanobert} employed pre-trained BERT for unsupervised learning with a masked language modeling loss function, and Hades utilized FastText for semantic vector representation and Transformer for classification.
(3) Performance disparities: previous empirical studies \cite{le2021log,le2022log} have indicated notably inferior performance of unsupervised and semi-supervised approaches compared to supervised approaches.

\subsection{Data Resampling for Software Engineering}
Data resampling has been widely applied to address the class imbalance issue in the field of software engineering, such as quality prediction \cite{seiffert2009improving},  bug classification \cite{zheng2021comparative, shu2021better}, defect prediction \cite{kamei2007effects, pelayo2012evaluating, tan2015online, bennin2017significant, huda2018ensemble, bennin2019relative, xu2021empirical, yedida2021value}, and code smell detection \cite{pecorelli2020large, li2023relative}. 
For example, Zheng et al. \cite{zheng2021comparative} analyzed the impact of six data resampling methods (e.g., SMOTE, Mahakil \cite{bennin2017mahakil}, and Rose \cite{lunardon2014rose}) on multiple classifiers for bug report classification. They found that the combination of Rose with random forest yielded the best performance. 
Bennin et al. \cite{bennin2017significant} observed that while their investigated data resampling methods have no statistically significant effect on defect prioritization, these methods improve the defect classification performance with regard to Recall and G-mean. Subsequently, they \cite{bennin2019relative} demonstrated that random undersampling and borderline-SMOTE are the more stable data resampling methods in software defect prediction.
Li et al. \cite{li2023relative} investigated the effects of 31 imbalanced learning methods on machine learning classifiers for code smell detection. Their study revealed varied impacts of these methods across different code smells, with deep forest consistently enhancing performance. Additionally, certain data resampling methods such as CNN, ENN, BSMOTE, and ROS showed superior performance compared to SMOTE.
Differently from prior work, in the context of log anomaly detection, which poses unique challenges such as the need for careful selection of window sizes, our research conducts an extensive analysis to explore the influence of data resampling methods on model performance across various window sizes. 
Furthermore, our study delves into the effects of some data resampling methods on both raw data and the feature space, providing valuable insights for practitioners in this field.

\subsection{Deep Learning-Based Anomaly Detection in Other Fields.}
Deep learning-based anomaly detection is applicable not only in software log anomaly detection but also in diverse fields like fraud detection \cite{heryadi2017learning, roy2018deep, raghavan2019fraud, hilal2022financial}, medical diagnosis \cite{tsiouris2018long, johnson2019survey, li2020classification, fernando2021deep}, manufacturing defect detection \cite{yang2020using}, and network intrusion detection \cite{abdelkhalek2023addressing}. While the workflow for anomaly detection in these domains shares similarities with the depicted Figure~\ref{fig:workflow}, specific data preprocessing steps, such as log parsing and grouping, may not be applicable. In these fields, class imbalance also presents a significant challenge that can affect the performance of anomaly detection models. Various data resampling methods have been employed to address this issue. 
For example, 
Roy et al. \cite{roy2018deep} implemented RUS with a recommended resampling ratio \cite{chen2008fast} of 10:1 for non-fraudulent to fraudulent credit card transaction data, followed by an LSTM model for detecting fraud in credit card transactions. 
Li et al. \cite{li2020classification} employed a CNN-based model with adjusted class weights to analyze phonocardiograms for abnormal heart sound detection. 
Abdelkhalek et al. \cite{abdelkhalek2023addressing} proposed a data resampling approach combining ADASYN and Tomek Links in conjunction with diverse deep learning models (such as LSTM and CNN) for improved detection of malicious attacks, aiming to address the class imbalance issue between normal traffic and attack samples. 
In summary, the aforementioned studies suggest that certain data resampling methods can enhance model performance, aligning with our findings. However, empirical research regarding the most effective data resampling methods, optimal resampling ratio, and their impacts on both raw data and the feature space is currently lacking in these fields.

\section{Conclusion}
Our study represents a pioneering effort in comprehensively assessing the impact of ten data resampling methods on alleviating class imbalance in DLLAD. 
Through empirical analysis, we have derived several critical insights. 
Firstly, severe data imbalances, like in the Thunderbird dataset, often lead to poor performance in DLLAD approaches. 
Secondly, oversampling methods generally outperform both undersampling and hybrid sampling methods. Moreover, data resampling on raw data yields superior results compared to data resampling in the feature space. Notably, \textit{ROS$_R$} exhibits outstanding performance, particularly in scenarios characterized by severe class imbalances.
Additionally, certain undersampling and hybrid sampling methods, such as \textit{SMOTEENN}, \textit{NearMiss}, and \textit{IHT}, show limited effectiveness in most cases. Our exploration of different resampling ratios of normal to abnormal data provides actionable recommendations for optimizing the impact of data resampling on DLLAD approaches. By adopting the recommended methods with specific resampling ratios, the performance of DLLAD approaches can be significantly enhanced. Furthermore, we offer implications for improved data resampling strategies through the oversampling of abnormal log sequences that contain important tokens contributing to DLLAD.

\textbf{Data Availability.} Our source-code and data are publicly available\footnote{https://github.com/ResamplingDLLAD/ResamplingEmpirical}.


\bibliographystyle{IEEEtran}
\bibliography{ref}

\begin{thebibliography}{10}
\providecommand{\url}[1]{#1}
\csname url@samestyle\endcsname
\providecommand{\newblock}{\relax}
\providecommand{\bibinfo}[2]{#2}
\providecommand{\BIBentrySTDinterwordspacing}{\spaceskip=0pt\relax}
\providecommand{\BIBentryALTinterwordstretchfactor}{4}
\providecommand{\BIBentryALTinterwordspacing}{\spaceskip=\fontdimen2\font plus
\BIBentryALTinterwordstretchfactor\fontdimen3\font minus \fontdimen4\font\relax}
\providecommand{\BIBforeignlanguage}[2]{{%
\expandafter\ifx\csname l@#1\endcsname\relax
\typeout{** WARNING: IEEEtran.bst: No hyphenation pattern has been}%
\typeout{** loaded for the language `#1'. Using the pattern for}%
\typeout{** the default language instead.}%
\else
\language=\csname l@#1\endcsname
\fi
#2}}
\providecommand{\BIBdecl}{\relax}
\BIBdecl

\bibitem{le2022log}
V.-H. Le and H.~Zhang, ``Log-based anomaly detection with deep learning: How far are we?'' in \emph{Proceedings of the 44th international conference on software engineering}, 2022, pp. 1356--1367.

\bibitem{le2021log}
{Le, Van-Hoang and Zhang, Hongyu}, ``Log-based anomaly detection without log parsing,'' in \emph{2021 36th IEEE/ACM International Conference on Automated Software Engineering (ASE)}.\hskip 1em plus 0.5em minus 0.4em\relax IEEE, 2021, pp. 492--504.

\bibitem{fu2013contextual}
Q.~Fu, J.-G. Lou, Q.~Lin, R.~Ding, D.~Zhang, and T.~Xie, ``Contextual analysis of program logs for understanding system behaviors,'' in \emph{2013 10th Working Conference on Mining Software Repositories (MSR)}.\hskip 1em plus 0.5em minus 0.4em\relax IEEE, 2013, pp. 397--400.

\bibitem{fu2014developers}
Q.~Fu, J.~Zhu, W.~Hu, J.-G. Lou, R.~Ding, Q.~Lin, D.~Zhang, and T.~Xie, ``Where do developers log? an empirical study on logging practices in industry,'' in \emph{Companion Proceedings of the 36th International Conference on Software Engineering}, 2014, pp. 24--33.

\bibitem{jiang2017causes}
H.~Jiang, X.~Li, Z.~Yang, and J.~Xuan, ``What causes my test alarm? automatic cause analysis for test alarms in system and integration testing,'' in \emph{2017 IEEE/ACM 39th International Conference on Software Engineering (ICSE)}.\hskip 1em plus 0.5em minus 0.4em\relax IEEE, 2017, pp. 712--723.

\bibitem{zhi2019exploratory}
C.~Zhi, J.~Yin, S.~Deng, M.~Ye, M.~Fu, and T.~Xie, ``An exploratory study of logging configuration practice in java,'' in \emph{2019 IEEE international conference on software maintenance and evolution (ICSME)}.\hskip 1em plus 0.5em minus 0.4em\relax IEEE, 2019, pp. 459--469.

\bibitem{he2021survey}
S.~He, P.~He, Z.~Chen, T.~Yang, Y.~Su, and M.~R. Lyu, ``A survey on automated log analysis for reliability engineering,'' \emph{ACM Computing Surveys}, vol.~54, no.~6, pp. 1--37, 2021.

\bibitem{du2017deeplog}
M.~Du, F.~Li, G.~Zheng, and V.~Srikumar, ``Deeplog: Anomaly detection and diagnosis from system logs through deep learning,'' in \emph{Proceedings of the 2017 ACM SIGSAC Conference on Computer and Communications Security (CCS)}, 2017, pp. 1285--1298.

\bibitem{lu2018detecting}
S.~Lu, X.~Wei, Y.~Li, and L.~Wang, ``Detecting anomaly in big data system logs using convolutional neural network,'' in \emph{2018 IEEE 16th Intl Conf on Dependable, Autonomic and Secure Computing, 16th Intl Conf on Pervasive Intelligence and Computing, 4th Intl Conf on Big Data Intelligence and Computing and Cyber Science and Technology Congress (DASC/PiCom/DataCom/CyberSciTech)}.\hskip 1em plus 0.5em minus 0.4em\relax IEEE, 2018, pp. 151--158.

\bibitem{zhang2019robust}
X.~Zhang, Y.~Xu, Q.~Lin, B.~Qiao, H.~Zhang, Y.~Dang, C.~Xie, X.~Yang, Q.~Cheng, Z.~Li \emph{et~al.}, ``Robust log-based anomaly detection on unstable log data,'' in \emph{Proceedings of the 2019 27th ACM Joint Meeting on European Software Engineering Conference and Symposium on the Foundations of Software Engineering (ESEC/SIGSOFT FSE)}, 2019, pp. 807--817.

\bibitem{meng2019loganomaly}
W.~Meng, Y.~Liu, Y.~Zhu, S.~Zhang, D.~Pei, Y.~Liu, Y.~Chen, R.~Zhang, S.~Tao, P.~Sun \emph{et~al.}, ``Loganomaly: Unsupervised detection of sequential and quantitative anomalies in unstructured logs.'' in \emph{Proceedings of the 28th International Joint Conference on Artificial Intelligence (IJCAI)}, 2019, pp. 4739--4745.

\bibitem{yang2021semi}
L.~Yang, J.~Chen, Z.~Wang, W.~Wang, J.~Jiang, X.~Dong, and W.~Zhang, ``Semi-supervised log-based anomaly detection via probabilistic label estimation,'' in \emph{2021 IEEE/ACM 43rd International Conference on Software Engineering (ICSE)}.\hskip 1em plus 0.5em minus 0.4em\relax IEEE, 2021, pp. 1448--1460.

\bibitem{ma2023semi}
X.~Ma, J.~Keung, P.~He, Y.~Xiao, X.~Yu, and Y.~Li, ``A semi-supervised approach for industrial anomaly detection via self-adaptive clustering,'' \emph{IEEE Transactions on Industrial Informatics}, 2023.

\bibitem{johnson2019survey}
J.~M. Johnson and T.~M. Khoshgoftaar, ``Survey on deep learning with class imbalance,'' \emph{Journal of Big Data}, vol.~6, no.~1, pp. 1--54, 2019.

\bibitem{van2007experimental}
J.~Van~Hulse, T.~M. Khoshgoftaar, and A.~Napolitano, ``Experimental perspectives on learning from imbalanced data,'' in \emph{Proceedings of the 24th international conference on Machine learning}, 2007, pp. 935--942.

\bibitem{chawla2002smote}
N.~V. Chawla, K.~W. Bowyer, L.~O. Hall, and W.~P. Kegelmeyer, ``{SMOTE}: synthetic minority over-sampling technique,'' \emph{Journal of artificial intelligence research}, vol.~16, pp. 321--357, 2002.

\bibitem{he2008adasyn}
H.~He, Y.~Bai, E.~A. Garcia, and S.~Li, ``{ADASYN}: Adaptive synthetic sampling approach for imbalanced learning,'' in \emph{2008 IEEE international joint conference on neural networks (IEEE world congress on computational intelligence)}.\hskip 1em plus 0.5em minus 0.4em\relax Ieee, 2008, pp. 1322--1328.

\bibitem{mani2003knn}
I.~Mani and I.~Zhang, ``{KNN} approach to unbalanced data distributions: a case study involving information extraction,'' in \emph{Proceedings of workshop on learning from imbalanced datasets}, vol. 126, no.~1.\hskip 1em plus 0.5em minus 0.4em\relax ICML, 2003, pp. 1--7.

\bibitem{smith2014instance}
M.~R. Smith, T.~Martinez, and C.~Giraud-Carrier, ``An instance level analysis of data complexity,'' \emph{Machine learning}, vol.~95, pp. 225--256, 2014.

\bibitem{batista2004study}
G.~E. Batista, R.~C. Prati, and M.~C. Monard, ``A study of the behavior of several methods for balancing machine learning training data,'' \emph{ACM SIGKDD explorations newsletter}, vol.~6, no.~1, pp. 20--29, 2004.

\bibitem{lin2016log}
Q.~Lin, H.~Zhang, J.-G. Lou, Y.~Zhang, and X.~Chen, ``Log clustering based problem identification for online service systems,'' in \emph{Proceedings of the 38th International Conference on Software Engineering Companion}, 2016, pp. 102--111.

\bibitem{he2017drain}
P.~He, J.~Zhu, Z.~Zheng, and M.~R. Lyu, ``Drain: An online log parsing approach with fixed depth tree,'' in \emph{2017 IEEE international conference on web services (ICWS)}.\hskip 1em plus 0.5em minus 0.4em\relax IEEE, 2017, pp. 33--40.

\bibitem{joulin2016fasttext}
A.~Joulin, E.~Grave, P.~Bojanowski, M.~Douze, H.~J{\'e}gou, and T.~Mikolov, ``Fasttext. zip: Compressing text classification models,'' \emph{arXiv preprint arXiv:1612.03651}, 2016.

\bibitem{salton_term-weighting_1988}
G.~Salton and C.~Buckley, ``Term-weighting approaches in automatic text retrieval,'' \emph{Information processing \& management}, vol.~24, no.~5, pp. 513--523, 1988.

\bibitem{devlin2018bert}
J.~Devlin, M.-W. Chang, K.~Lee, and K.~Toutanova, ``Bert: Pre-training of deep bidirectional transformers for language understanding,'' \emph{arXiv preprint arXiv:1810.04805}, 2018.

\bibitem{chakraborty2021deep}
S.~Chakraborty, R.~Krishna, Y.~Ding, and B.~Ray, ``Deep learning based vulnerability detection: Are we there yet,'' \emph{IEEE Transactions on Software Engineering}, 2021.

\bibitem{yang2023does}
X.~Yang, S.~Wang, Y.~Li, and S.~Wang, ``Does data sampling improve deep learning-based vulnerability detection? yeas! and nays!'' in \emph{2023 IEEE/ACM 45th International Conference on Software Engineering (ICSE)}.\hskip 1em plus 0.5em minus 0.4em\relax IEEE, 2023, pp. 2287--2298.

\bibitem{pelayo2012evaluating}
L.~Pelayo and S.~Dick, ``Evaluating stratification alternatives to improve software defect prediction,'' \emph{IEEE transactions on reliability}, vol.~61, no.~2, pp. 516--525, 2012.

\bibitem{malhotra2019empirical}
R.~Malhotra and S.~Kamal, ``An empirical study to investigate oversampling methods for improving software defect prediction using imbalanced data,'' \emph{Neurocomputing}, vol. 343, pp. 120--140, 2019.

\bibitem{breiman2001random}
L.~Breiman, ``Random forests,'' \emph{Machine learning}, vol.~45, pp. 5--32, 2001.

\bibitem{wilson1972asymptotic}
D.~L. Wilson, ``Asymptotic properties of nearest neighbor rules using edited data,'' \emph{IEEE Transactions on Systems, Man, and Cybernetics}, no.~3, pp. 408--421, 1972.

\bibitem{tomek1976two}
I.~Tomek, ``Two modifications of {CNN}.'' 1976.

\bibitem{he2020loghub}
S.~He, J.~Zhu, P.~He, and M.~R. Lyu, ``Loghub: A large collection of system log datasets towards automated log analytics,'' \emph{arXiv preprint arXiv:2008.06448}, 2020.

\bibitem{oliner2007supercomputers}
A.~Oliner and J.~Stearley, ``What supercomputers say: A study of five system logs,'' in \emph{37th annual IEEE/IFIP international conference on dependable systems and networks (DSN'07)}.\hskip 1em plus 0.5em minus 0.4em\relax IEEE, 2007, pp. 575--584.

\bibitem{lee2023heterogeneous}
C.~Lee, T.~Yang, Z.~Chen, Y.~Su, Y.~Yang, and M.~R. Lyu, ``Heterogeneous anomaly detection for software systems via semi-supervised cross-modal attention,'' in \emph{2023 IEEE/ACM 45th International Conference on Software Engineering (ICSE)}, 2023.

\bibitem{liu2021lognads}
X.~Liu, W.~Liu, X.~Di, J.~Li, B.~Cai, W.~Ren, and H.~Yang, ``Lognads: Network anomaly detection scheme based on log semantics representation,'' \emph{Future Generation Computer Systems}, vol. 124, pp. 390--405, 2021.

\bibitem{lee2023lanobert}
Y.~Lee, J.~Kim, and P.~Kang, ``Lanobert: System log anomaly detection based on bert masked language model,'' \emph{Applied Soft Computing}, vol. 146, p. 110689, 2023.

\bibitem{he2016experience}
S.~He, J.~Zhu, P.~He, and M.~R. Lyu, ``Experience report: System log analysis for anomaly detection,'' in \emph{2016 IEEE 27th international symposium on software reliability engineering (ISSRE)}.\hskip 1em plus 0.5em minus 0.4em\relax IEEE, 2016, pp. 207--218.

\bibitem{song2018comprehensive}
Q.~Song, Y.~Guo, and M.~Shepperd, ``A comprehensive investigation of the role of imbalanced learning for software defect prediction,'' \emph{IEEE Transactions on Software Engineering}, vol.~45, no.~12, pp. 1253--1269, 2018.

\bibitem{yao2020assessing}
J.~Yao and M.~Shepperd, ``Assessing software defection prediction performance: Why using the matthews correlation coefficient matters,'' in \emph{Proceedings of the 24th International Conference on Evaluation and Assessment in Software Engineering}, 2020, pp. 120--129.

\bibitem{moussa2022use}
R.~Moussa and F.~Sarro, ``On the use of evaluation measures for defect prediction studies,'' in \emph{Proceedings of the 31st ACM SIGSOFT International Symposium on Software Testing and Analysis}, 2022, pp. 101--113.

\bibitem{bennin2019relative}
K.~E. Bennin, J.~W. Keung, and A.~Monden, ``On the relative value of data resampling approaches for software defect prediction,'' \emph{Empirical Software Engineering}, vol.~24, pp. 602--636, 2019.

\bibitem{bennin2022empirical}
K.~E. Bennin, A.~Tahir, S.~G. MacDonell, and J.~B{\"o}rstler, ``An empirical study on the effectiveness of data resampling approaches for cross-project software defect prediction,'' \emph{IET Software}, vol.~16, no.~2, pp. 185--199, 2022.

\bibitem{tantithamthavorn2016empirical}
C.~Tantithamthavorn, S.~McIntosh, A.~E. Hassan, and K.~Matsumoto, ``An empirical comparison of model validation techniques for defect prediction models,'' \emph{IEEE Transactions on Software Engineering}, vol.~43, no.~1, pp. 1--18, 2016.

\bibitem{lemaitre2017imbalanced}
G.~Lema{\^\i}tre, F.~Nogueira, and C.~K. Aridas, ``Imbalanced-learn: A python toolbox to tackle the curse of imbalanced datasets in machine learning,'' \emph{The Journal of Machine Learning Research}, vol.~18, no.~1, pp. 559--563, 2017.

\bibitem{yu2024deep}
B.~Yu, J.~Yao, Q.~Fu, Z.~Zhong, H.~Xie, Y.~Wu, Y.~Ma, and P.~He, ``Deep learning or classical machine learning? an empirical study on log-based anomaly detection,'' in \emph{Proceedings of the 46th IEEE/ACM International Conference on Software Engineering}, 2024, pp. 1--13.

\bibitem{ribeiro2016should}
M.~T. Ribeiro, S.~Singh, and C.~Guestrin, ``" why should i trust you?" explaining the predictions of any classifier,'' in \emph{Proceedings of the 22nd ACM SIGKDD international conference on knowledge discovery and data mining}, 2016, pp. 1135--1144.

\bibitem{pornprasit2021pyexplainer}
C.~Pornprasit, C.~Tantithamthavorn, J.~Jiarpakdee, M.~Fu, and P.~Thongtanunam, ``Pyexplainer: Explaining the predictions of just-in-time defect models,'' in \emph{2021 36th IEEE/ACM International Conference on Automated Software Engineering (ASE)}.\hskip 1em plus 0.5em minus 0.4em\relax IEEE, 2021, pp. 407--418.

\bibitem{tantithamthavorn2021explainable}
C.~K. Tantithamthavorn and J.~Jiarpakdee, ``Explainable ai for software engineering,'' in \emph{2021 36th IEEE/ACM International Conference on Automated Software Engineering (ASE)}.\hskip 1em plus 0.5em minus 0.4em\relax IEEE, 2021, pp. 1--2.

\bibitem{ledel2022studying}
B.~Ledel and S.~Herbold, ``Studying the explanations for the automated prediction of bug and non-bug issues using lime and shap,'' \emph{arXiv preprint arXiv:2209.07623}, 2022.

\bibitem{fan2020can}
M.~Fan, W.~Wei, X.~Xie, Y.~Liu, X.~Guan, and T.~Liu, ``Can we trust your explanations? sanity checks for interpreters in android malware analysis,'' \emph{IEEE Transactions on Information Forensics and Security}, vol.~16, pp. 838--853, 2020.

\bibitem{feichtner2020understanding}
J.~Feichtner and S.~Gruber, ``Understanding privacy awareness in android app descriptions using deep learning,'' in \emph{Proceedings of the tenth ACM conference on data and application security and privacy}, 2020, pp. 203--214.

\bibitem{warnecke2020evaluating}
A.~Warnecke, D.~Arp, C.~Wressnegger, and K.~Rieck, ``Evaluating explanation methods for deep learning in security,'' in \emph{2020 IEEE european symposium on security and privacy (EuroS\&P)}.\hskip 1em plus 0.5em minus 0.4em\relax IEEE, 2020, pp. 158--174.

\bibitem{lyu2021towards}
Y.~Lyu, G.~K. Rajbahadur, D.~Lin, B.~Chen, and Z.~M. Jiang, ``Towards a consistent interpretation of aiops models,'' \emph{ACM Transactions on Software Engineering and Methodology (TOSEM)}, vol.~31, no.~1, pp. 1--38, 2021.

\bibitem{steenhoek2023empirical}
B.~Steenhoek, M.~M. Rahman, R.~Jiles, and W.~Le, ``An empirical study of deep learning models for vulnerability detection,'' in \emph{2023 IEEE/ACM 45th International Conference on Software Engineering (ICSE)}.\hskip 1em plus 0.5em minus 0.4em\relax IEEE, 2023, pp. 2237--2248.

\bibitem{zhang2016automated}
K.~Zhang, J.~Xu, M.~R. Min, G.~Jiang, K.~Pelechrinis, and H.~Zhang, ``Automated it system failure prediction: A deep learning approach,'' in \emph{2016 IEEE International Conference on Big Data (Big Data)}.\hskip 1em plus 0.5em minus 0.4em\relax IEEE, 2016, pp. 1291--1300.

\bibitem{hochreiter1997long}
S.~Hochreiter and J.~Schmidhuber, ``Long short-term memory,'' \emph{Neural computation}, vol.~9, no.~8, pp. 1735--1780, 1997.

\bibitem{vinayakumar2017long}
R.~Vinayakumar, K.~Soman, and P.~Poornachandran, ``Long short-term memory based operation log anomaly detection,'' in \emph{2017 International Conference on Advances in Computing, Communications and Informatics (ICACCI)}.\hskip 1em plus 0.5em minus 0.4em\relax IEEE, 2017, pp. 236--242.

\bibitem{li_swisslog_2020}
X.~Li, P.~Chen, L.~Jing, Z.~He, and G.~Yu, ``Swisslog: Robust and unified deep learning based log anomaly detection for diverse faults,'' in \emph{2020 IEEE 31st International Symposium on Software Reliability Engineering (ISSRE)}.\hskip 1em plus 0.5em minus 0.4em\relax IEEE, 2020, pp. 92--103.

\bibitem{seiffert2009improving}
C.~Seiffert, T.~M. Khoshgoftaar, and J.~Van~Hulse, ``Improving software-quality predictions with data sampling and boosting,'' \emph{IEEE Transactions on Systems, Man, and Cybernetics-Part A: Systems and Humans}, vol.~39, no.~6, pp. 1283--1294, 2009.

\bibitem{zheng2021comparative}
W.~Zheng, Y.~Xun, X.~Wu, Z.~Deng, X.~Chen, and Y.~Sui, ``A comparative study of class rebalancing methods for security bug report classification,'' \emph{IEEE Transactions on Reliability}, vol.~70, no.~4, pp. 1658--1670, 2021.

\bibitem{shu2021better}
R.~Shu, T.~Xia, J.~Chen, L.~Williams, and T.~Menzies, ``How to better distinguish security bug reports (using dual hyperparameter optimization),'' \emph{Empirical Software Engineering}, vol.~26, pp. 1--37, 2021.

\bibitem{kamei2007effects}
Y.~Kamei, A.~Monden, S.~Matsumoto, T.~Kakimoto, and K.-i. Matsumoto, ``The effects of over and under sampling on fault-prone module detection,'' in \emph{First international symposium on empirical software engineering and measurement (ESEM 2007)}.\hskip 1em plus 0.5em minus 0.4em\relax IEEE, 2007, pp. 196--204.

\bibitem{tan2015online}
M.~Tan, L.~Tan, S.~Dara, and C.~Mayeux, ``Online defect prediction for imbalanced data,'' in \emph{2015 IEEE/ACM 37th IEEE International Conference on Software Engineering}, vol.~2.\hskip 1em plus 0.5em minus 0.4em\relax IEEE, 2015, pp. 99--108.

\bibitem{bennin2017significant}
K.~E. Bennin, J.~Keung, A.~Monden, P.~Phannachitta, and S.~Mensah, ``The significant effects of data sampling approaches on software defect prioritization and classification,'' in \emph{2017 ACM/IEEE International Symposium on Empirical Software Engineering and Measurement (ESEM)}.\hskip 1em plus 0.5em minus 0.4em\relax IEEE, 2017, pp. 364--373.

\bibitem{huda2018ensemble}
S.~Huda, K.~Liu, M.~Abdelrazek, A.~Ibrahim, S.~Alyahya, H.~Al-Dossari, and S.~Ahmad, ``An ensemble oversampling model for class imbalance problem in software defect prediction,'' \emph{IEEE access}, vol.~6, pp. 24\,184--24\,195, 2018.

\bibitem{xu2021empirical}
H.~Xu, R.~Duan, S.~Yang, and L.~Guo, ``An empirical study on data sampling for just-in-time defect prediction,'' in \emph{Artificial Intelligence and Security: 7th International Conference, ICAIS 2021, Dublin, Ireland, July 19--23, 2021, Proceedings, Part II 7}.\hskip 1em plus 0.5em minus 0.4em\relax Springer, 2021, pp. 54--69.

\bibitem{yedida2021value}
R.~Yedida and T.~Menzies, ``On the value of oversampling for deep learning in software defect prediction,'' \emph{IEEE Transactions on Software Engineering}, vol.~48, no.~8, pp. 3103--3116, 2021.

\bibitem{pecorelli2020large}
F.~Pecorelli, D.~Di~Nucci, C.~De~Roover, and A.~De~Lucia, ``A large empirical assessment of the role of data balancing in machine-learning-based code smell detection,'' \emph{Journal of Systems and Software}, vol. 169, p. 110693, 2020.

\bibitem{li2023relative}
F.~Li, K.~Zou, J.~W. Keung, X.~Yu, S.~Feng, and Y.~Xiao, ``On the relative value of imbalanced learning for code smell detection,'' \emph{Software: Practice and Experience}, 2023.

\bibitem{bennin2017mahakil}
K.~E. Bennin, J.~Keung, P.~Phannachitta, A.~Monden, and S.~Mensah, ``Mahakil: Diversity based oversampling approach to alleviate the class imbalance issue in software defect prediction,'' \emph{IEEE Transactions on Software Engineering}, vol.~44, no.~6, pp. 534--550, 2017.

\bibitem{lunardon2014rose}
N.~Lunardon, G.~Menardi, and N.~Torelli, ``Rose: a package for binary imbalanced learning.'' \emph{R journal}, vol.~6, no.~1, 2014.

\bibitem{heryadi2017learning}
Y.~Heryadi and H.~L. H.~S. Warnars, ``Learning temporal representation of transaction amount for fraudulent transaction recognition using {CNN}, stacked {LSTM}, and {CNN-LSTM},'' in \emph{2017 IEEE International Conference on Cybernetics and Computational Intelligence (CyberneticsCom)}.\hskip 1em plus 0.5em minus 0.4em\relax IEEE, 2017, pp. 84--89.

\bibitem{roy2018deep}
A.~Roy, J.~Sun, R.~Mahoney, L.~Alonzi, S.~Adams, and P.~Beling, ``Deep learning detecting fraud in credit card transactions,'' in \emph{2018 systems and information engineering design symposium (SIEDS)}.\hskip 1em plus 0.5em minus 0.4em\relax IEEE, 2018, pp. 129--134.

\bibitem{raghavan2019fraud}
P.~Raghavan and N.~El~Gayar, ``Fraud detection using machine learning and deep learning,'' in \emph{2019 international conference on computational intelligence and knowledge economy (ICCIKE)}.\hskip 1em plus 0.5em minus 0.4em\relax IEEE, 2019, pp. 334--339.

\bibitem{hilal2022financial}
W.~Hilal, S.~A. Gadsden, and J.~Yawney, ``Financial fraud: a review of anomaly detection techniques and recent advances,'' \emph{Expert systems With applications}, vol. 193, p. 116429, 2022.

\bibitem{tsiouris2018long}
K.~M. Tsiouris, V.~C. Pezoulas, M.~Zervakis, S.~Konitsiotis, D.~D. Koutsouris, and D.~I. Fotiadis, ``A long short-term memory deep learning network for the prediction of epileptic seizures using eeg signals,'' \emph{Computers in biology and medicine}, vol.~99, pp. 24--37, 2018.

\bibitem{li2020classification}
F.~Li, H.~Tang, S.~Shang, K.~Mathiak, and F.~Cong, ``Classification of heart sounds using convolutional neural network,'' \emph{Applied Sciences}, vol.~10, no.~11, p. 3956, 2020.

\bibitem{fernando2021deep}
T.~Fernando, H.~Gammulle, S.~Denman, S.~Sridharan, and C.~Fookes, ``Deep learning for medical anomaly detection--a survey,'' \emph{ACM Computing Surveys (CSUR)}, vol.~54, no.~7, pp. 1--37, 2021.

\bibitem{yang2020using}
J.~Yang, S.~Li, Z.~Wang, H.~Dong, J.~Wang, and S.~Tang, ``Using deep learning to detect defects in manufacturing: a comprehensive survey and current challenges,'' \emph{Materials}, vol.~13, no.~24, p. 5755, 2020.

\bibitem{abdelkhalek2023addressing}
A.~Abdelkhalek and M.~Mashaly, ``Addressing the class imbalance problem in network intrusion detection systems using data resampling and deep learning,'' \emph{The Journal of Supercomputing}, pp. 1--34, 2023.

\bibitem{chen2008fast}
X.-w. Chen and M.~Wasikowski, ``Fast: a roc-based feature selection metric for small samples and imbalanced data classification problems,'' in \emph{Proceedings of the 14th ACM SIGKDD international conference on Knowledge discovery and data mining}, 2008, pp. 124--132.

\end{thebibliography}

\newpage

 




\vfill

\end{document}